\documentclass[12pt]{article}

\usepackage{graphicx}
\usepackage{mathrsfs}
\usepackage{latexsym}
\usepackage{amsmath}
\usepackage{amssymb}
\usepackage{color}
\usepackage{subfig}
\usepackage{array}
\usepackage{multirow}
\usepackage{hhline}
\usepackage[table]{xcolor}
\usepackage{authblk}

\usepackage[nohead,letterpaper]{geometry}
\usepackage{vmargin}
\setpapersize{USletter}
\setmargrb{2cm}{1cm}{2cm}{1.75cm}

\usepackage{xcolor}
\definecolor{plum}{rgb}{0.36078, 0.20784, 0.4}
\definecolor{chameleon}{rgb}{0.30588, 0.60392, 0.023529}
\definecolor{cornflower}{rgb}{0.12549, 0.29020, 0.52941}
\definecolor{scarlet}{rgb}{0.8, 0, 0}
\definecolor{brick}{rgb}{0.64314, 0, 0}

\newcommand{\email}[1]{\href{mailto:#1}{\tt \textcolor{cornflower}{#1}}}

\newcommand{\ba}{\begin{eqnarray}}
\newcommand{\ea}{\end{eqnarray}}
\newcommand{\be}{\begin{equation}}
\newcommand{\ee}{\end{equation}}
\newcommand{\bd}{\begin{displaymath}}
\newcommand{\ed}{\end{displaymath}}

\usepackage[%
  pdftitle={},%
  pdfauthor={Miok Park},%
  pdfsubject={},%
  pdfkeywords={},%
  colorlinks=true,linkcolor=cornflower,citecolor=brick,urlcolor=chameleon%
  ]{hyperref}

\numberwithin{equation}{section}

\usepackage[bf,medium,]{titlesec}

\titleformat{\section}
{\large\bfseries}{\thesection.}{.5em}{\large\bfseries}

\titleformat{\subsection}
{\bfseries}{\thesubsection}{.5em}{\bfseries}

\newcommand{\cmb}[1]{{\color{blue}{#1}}} 

\begin{document}

\title{Complement to thermodynamics of dyonic Taub-NUT-AdS spacetime}

\date{}

\vspace{1cm}

\author[a]{Robert B. Mann}
\author[b,c]{Leopoldo A. Pando Zayas}
\author[d]{Miok Park}

\affil[a]{\it{\small{Department of Physics and Astronomy, University of Waterloo, Waterloo, Ontario, Canada, N2L 3G1}}}
\affil[b]{\it{\small{Leinweber Center for Theoretical Physics, University of Michigan, Ann Arbor, MI 48109, U.S.A.}}}
\affil[c]{\it{\small{The Abdus Salam International Centre for Theoretical Physics, 34014 Trieste, Italy}}}
\affil[d]{\it{\small{School of Physics, Korea Institute for Advanced Study, Seoul 02455, Republic of Korea}}}

\maketitle
\center{
\email{$^{a}$rbmann@uwaterloo.ca, $^{b,c}$lpandoz@umich.edu, $^d$miokpark@kias.re.kr,}}

\vspace{1cm}
\abstract{We examine the thermodynamics of Euclidean dyonic Taub-NUT/Bolt-AdS$_4$ black holes for a variety of horizon geometries to understand how gauge field regularity conditions   influence the thermodynamic relations. We find several distinct features that distinguish the NUT-charged case from its  dyonic Reissner-Nordstrom counterpart.  For the 
 NUT  solution,  the gauge field vanishes at the horizon  and so   regularity is ensured. For the Bolt solution we find that the norm of the gauge field is required to 
 vanish at the horizon in order to satisfy both regularity and the first law of thermodynamics. This regularity condition yields a constraint on the electric and magnetic charges and so reduces cohomogeneity of the system;  for spherical horizons, the regularity condition removing the Misner string singularity  further reduces cohomogeneity,  We observe that   bolt solutions  with increasing electric charge  have positive heat capacity,  but upon  turning on the magnetic charge to make
 the solution dyonic, we find that the properties of the uncharged one are retained, having both positive and negative heat capacity.  We also study 
  the extremal Bolt solution, finding  that Misner string disappears  at the horizon in the zero temperature limit.  We find that the extremal solution has 
   finite-temperature-like behaviour, with   the electric potential playing a role similar to  temperature.}

\newpage
\tableofcontents

\newpage


\section{Introduction}

One of  the more interesting solutions to the Einstein equations is Taub-NUT spacetime,  a generalization of the Schwarzschild spacetime originally obtained by  Taub in 1951 \cite{Taub:1950ez}  and later rediscovered by  Newman,  Tamburino, and  Unti in 1963 \cite{Newman:1963yy}. This spacetime is asymptotically locally flat: its Riemann tensor approaches zero at large distance, but  it is only locally isomorphic to Minkowski spacetime in this regime.  Its (anti) de Sitter ((A)dS) version,  Taub-NUT-(A)dS spacetime, is likewise asymptotically locally (A)dS: its Riemann tensor 
 \begin{equation}
R_{\alpha \beta \mu \nu} \rightarrow \mp \frac{1}{l^2} \bigg(g_{\alpha \mu} g_{\beta \nu} - g_{\alpha \nu} g_{\beta \mu} \bigg)
\end{equation} 
at large distance where $l$ is the (A)dS$_4$ length. These Taub-NUT(-(A)dS) spacetimes also have a mass parameter ($m$) and a NUT charge ($s$), with horizons at $g_{tt}(r_+)=0$. However the Kretschmann scalar does not diverge at the origin $r=0$ unless $s=0$.

Interestingly, these spacetimes have peculiar properties when their horizon geometry is a sphere due to the non-trivial fibration of time on $S^2$, which induces spurious singularity in a given coordinate system. Known as a Misner string   \cite{Misner:1963fr}, this singularity is  analogous to the Dirac string of electromagnetism, and the NUT charge is a gravitomagnetic charge analogous to a  magnetic monopole.  Hence any Taub-NUT solution with nonzero $m$ and $s$ is a gravitational dyon. Like the Dirac string, the Misner string that appears along the $\theta = \pi$ (or $\theta =0$) axis can be eliminated 
  by introducing an additional time coordinate in a neighbourhood $\theta = \pi$ (or $\theta =0$) that is regular there, and then gluing this region to its complement; a similar construction can be made for a Misner string  along any axis.  However this regularity condition automatically requires the  time coordinate to be periodic $t \sim t + 8 \pi s$, which yields  closed timelike curves (CTCs) through every point. While this is a benign situation in Euclidean space, it violates causality in Lorentzian spacetime, and so the latter has generally been thought to be   pathological.
  
Since the Taub-NUT-((A)dS) spacetime pertains these peculiar features, understanding their thermodynamic properties were challenging, mainly due to the violation of Bekenstein-Hawking formula, but successfully studied in \cite{Hawking:1998jf,Hawking:1998ct,Mann:1999pc}. Moreover, as this solution can be inherited from M-theory or supergravity theory, its higher dimensional solutions are studied in \cite{Taylor:1998fd,Clarkson:2002uj} and its supersymmetric properties are found in \cite{AlonsoAlberca:2000cs}. Also the asymptotically locally AdS spacetime can be considered as a natural extension of AdS/CFT correspondence and so holographic matching with its boundary field theory are studied in \cite{Martelli:2012sz,Toldo:2017qsh}.

Recently a new perspective on these spacetimes has emerged, based on studies \cite{Clement:2015cxa}  showing that the Misner string in Lorentzian Taub-NUT spacetime is transparent for geodesics, implying that  the spacetime becomes geodesically complete. Thus imposing time periodicity is not necessary and no causality problem occurs. Based on this argument, thermodynamic properties of Lorentzian Taub-NUT-AdS spacetime were subsequently studied \cite{Kubiznak:2019yiu,Ballon:2019uha,Bordo:2020kxm}, with the perspective that the  first law has full cohomogeneity by treating the NUT charge as an independent thermodynamic variable, a Misner charge whose conjugate variable is called the Misner potential, with these new terms  included in the first law and free energy.

Here we consider Euclidean dyonic Taub-NUT-AdS$_4$ spacetime and investigate its thermodynamic properties for all horizon geometries characterized by  a parameter $\kappa$, where $\kappa=1$ corresponds to spherical geometries, $\kappa=0$ to locally flat geometries, and $\kappa < 0$ to hyperbolic geometries. We also consider the role played by the Misner string if $\kappa=1$. This extends previous studies \cite{Chamblin:1998pz,Johnson:2014xza,Johnson:2014pwa}; our aim here is to understand how regularity conditions of the gauge field affect  the thermodynamic relations.   For the dyonic RN black hole, the first law of thermodynamics is not strongly affected by the regularity condition of the gauge field, and an electric and magnetic charge are independently present in the thermodynamic relations as illustrated in appendix \ref{sec:appRNbh}. However  for the dyonic Taub-NUT-AdS solution,  we find that the regularity condition, which relates the electric and magnetic charges, is essential for satisfying the first law of thermodynamics. 

To start with, we perform the Wick-rotation $t \rightarrow - i \tau$ and $s \rightarrow - i \chi$ to Euclidean space from the Lorentzian solution. The $U(1)$ isometry group associated with the Killing vector $\xi = \partial_{\tau}$ then has either a zero-dimensional fixed point set (the so-called nut solution, or TN-AdS) or a  two-dimensional fixed point set (the so-called bolt solution, or TB-AdS). When the horizon geometry is that of a sphere ($\kappa=1$) it is of special interest because of the existence of the Misner string, which is removed by imposing the time periodicity as stated above. The Hawking temperature  is then  constrained to be $\frac{1}{8 \pi \chi}$, and the  Misner string contributes to the entropy. Thus the entropy is not just the area of the fixed point set of $\xi$,   which is the Bekenstein-Hawking entropy formula \cite{Hawking:1998ct,Mann:1999pc}. With these considerations, the thermodynamic properties of (a charged) Taub-NUT-AdS spacetime for $\kappa=1$  have been studied \cite{Chamblin:1998pz,Johnson:2014xza,Johnson:2014pwa}; for the uncharged 
TB-AdS case  there is a maximum value of the NUT charge. Each value of NUT charge has two branches \cite{Chamblin:1998pz}, characterized by increasing/decreasing horizon radius  as the NUT charge increases to its maximum value. We shall refer to them as small and  large TB-AdS solutions. They respectively have   negative and   positive heat capacities.   

The TN-AdS solution  demands that $P=i Q$, which in turn implies  the norm of the gauge field vanishes at the horizon.  However for TB-AdS, the zero value of this norm must be imposed (rather than having a finite value) to ensure the validity of the thermodynamic relations.  One way of   satisfying this regularity condition for TB-AdS is to relate electric and  magnetic charges: 
\begin{equation}
P = \frac{2 i Q \chi r_+}{r_+^2 + \chi^2}
\end{equation}
implying $Q$ and $P$ are no longer independent. This reduces the cohomogeneity of the system. Removal  of the Misner string if $\kappa=1$ yields a condition that further reduces the cohomogeneity in the thermodynamic relations. In this case the following thermodynamic relations 
\begin{align}
dE = T dS + \Phi_E dQ, \qquad F = E - TS - \Phi_E Q 
\end{align}
are satisfied. 
 
 We also  find that these thermodynamic arguments are valid in the zero temperature limit and study the near horizon geometry, which is   $AdS_2 \times \mathcal{M}_{\kappa}$. For $\kappa=1$, it is possible to make sense of the thermodynamics provided the temperature is not identified via Misner string periodicity arguments. We  observe that, upon replacing the temperature with the electric potential \cite{Choi:2018vbz}, the extremal TB-AdS solution exhibits similar thermodynamic behaviour to that of the uncharged TB-AdS solution.

Our manuscript is organized as follows. In section 2, we  review the thermodynamics for the uncharged TN/TB-AdS.  We shall work at fixed AdS length, not working in the extended thermodynamic phase space \cite{Kubiznak:2016qmn}. We check the first law of black hole thermodynamics and free energy and plotted the horizon of TB-AdS, entropy and heat capacity as functions of the NUT charge $\chi$, which by regularity requirements is the inverse of temperature for $\kappa=1$. In section 3, we revisit the thermodynamics of the charged TN-AdS and TB-AdS. We recover some results obtained previously \cite{Johnson:2014pwa}, but go beyond this by considering all values of  $\kappa$, calculating the heat capacity and other relevant thermodynamic quantities before imposing the regularity condition.  We find that   $P$ and $Q$ are related by metric requirements for TN-AdS and by the regularity condition of the gauge field at the horizon of TB-AdS.    We impose the regularity condition after obtaining the various thermodynamic quantities to show that applying the regularity condition is necessary for the thermodynamic relations to hold.  In section 4, we then investigate the near horizon geometry in the zero temperature limit, and study the resultant thermodynamic relations. Then we summarize our results in section 5.

\section{Recap of thermodynamics of Uncharged TN/TB-AdS}

The Einstein-Hilbert action with cosmological constant and Gibbons-Hawking term in four dimensional spacetime is
\begin{equation}\label{fullact}
I=  \frac{1}{16 \pi G_4} \int_{\cal M} d^4 x \sqrt{-g} \bigg(R + \frac{6}{l^2} \bigg) + \frac{1}{8 \pi G_4} \int_{\partial\cal M} d^3 x \sqrt{-h} K \cmb{-} \frac{1}{8 \pi G_4} \int_{\partial\cal M} d^3 x \sqrt{-h} \bigg( \frac{2}{l} + \frac{l}{2} R_3(h) \bigg)
\end{equation}
where $l$ is the radius of $AdS_4$ spacetime,  $K$ is the trace of extrinsic curvature on the boundary at spatial infinity, and the last term is
the boundary counterterm action $I_{ct}$ \cite{Mann:1999pc, Balasubramanian:1999re, Emparan:1999pm}.

 The variation of the action with respect to the metric yields the equations of motion
\begin{align}
&R_{\mu \nu} - \frac{1}{2} g_{\mu \nu} R - \frac{3}{l^2} g_{\mu \nu} = 0
\end{align}
with the solution
\begin{equation}\label{tnbh}
ds^2 = -f(r)(d t + 2 s \lambda(\theta) d \phi)^2 + \frac{dr^2}{f(r)} + (r^2 + s^2) (d \theta^2 + Y(\theta)^2 d \phi^2)
\end{equation}
where $s$ is the NUT charge,
\begin{equation}
f(r) = \frac{l^{-2}(r^2 + s^2)^2 + (\kappa + 4 l^{-2} s^2)(r^2 - s^2) - 2 M r}{r^2 + s^2}
\end{equation}
and
\begin{align}\label{mankappa}
\lambda(\theta) = 
  \begin{cases}
    \; \; \; \; \cos \theta     \\
    \; \; \; \; \; - \theta \\
    \; \; -\cosh \theta  \\
  \end{cases}, \qquad
Y(\theta) = 
  \begin{cases}
    \; \; \sin \theta      & \quad \text{for } \kappa = 1 \\
    \; \; \; \; 1 & \quad \text{for } \kappa = 0 \\
    \; \; \sinh \theta  & \quad \text{for } \kappa = -1. \\
  \end{cases}
\end{align}
characterize the three possible geometries of the constant-curvature horizon, located at the largest root  of $f(r)=0$.

To explore the thermodynamic properties of the spacetime \eqref{tnbh}, we shall compute its free-energy under various circumstances.  Performing the
  Wick-rotation $t \rightarrow - i \tau$ and $s \rightarrow - i \chi$, the Euclidean action and associated metric are
\begin{align}
I_E &= - \frac{1}{16 \pi G_4} \int d^4 x \sqrt{g} \bigg(R + \frac{6}{l^2}  \bigg) - \frac{1}{8 \pi G_4} \int d^3 x \sqrt{h} \bigg(K -\frac{2}{l} - \frac{l}{2} R_3 \bigg) \label{eq:Iepq0}\\
ds_{\textrm{E}}^2 &= f_E(r)(d \tau + 2 \chi \lambda(\theta) d \phi)^2 + \frac{dr^2}{f_E(r)} + (r^2 - \chi^2) (d \theta^2 + Y(\theta)^2 d \phi^2), \label{eq:metricPQ0} 
 \\
f_E &= \frac{l^{-2}(r^2 - \chi^2)^2 + (\kappa - 4 l^{-2} \chi^2)(r^2 + \chi^2) - 2 M r}{r^2 - \chi^2}  
\label{eucf}
\end{align}
This solution has an horizon at $r=r_+$ from $f_E(r_+) = 0$.  There are two qualitatively distinct solutions that are characterized by  the fixed point sets associated with the Killing vector field $\xi = \partial_{\tau}$.  If the radius is the same as the NUT charge $(r_+=\chi)$, the fixed point set is zero-dimensional, and is referred to  as the  Taub-Nut-AdS (TN-AdS) solution. Otherwise $r_+ > \chi$, in which case the fixed point sets is two-dimensional and is called the Taub-Bolt-AdS(TB-AdS) solution.

\subsection{Uncharged TN-AdS}

The TN-AdS solution is obtained when   $f_{\textrm{E}}(r)$ has a root at  $r=r_+ = \chi$.  The Euclidean metric function then can be written as
\begin{align}
f_n(r) = &\frac{(r-\chi ) \left(\kappa  l^2+r^2+2 r \chi -3 \chi ^2\right)}{l^2 (r+\chi )}
\end{align}
where in \eqref{eucf}
\begin{equation}
M  = M_n = \chi  \kappa  - \frac{4 \chi^3}{l^2}
\end{equation}
since $r_+=\chi$. The
Hawking temperature is easily calculated as 
\begin{equation}
T_n=\frac{1}{4 \pi} f'_n(\chi) = \frac{\kappa}{8 \pi \chi}
\end{equation}
and this automatically agrees with the condition for removal of the Misner string.When $\kappa = -1$, $f_n$ becomes negative near the horizon limit $r = \chi + \epsilon$.  This happens because   $r=\chi$ is not the outermost horizon, but instead is contained in a bolt solution whose horizon is $\sqrt{4 \chi ^2-\kappa  l^2}-\chi$ \cite{Mann:2004mi}. Thus there is no hyperbolic TN-AdS solution. We therefore only consider the $\kappa=0$ and $\kappa=1$ cases for TN-AdS, noting that the former is a zero-temperature extremal case.

The extrinsic curvature and the three dimensional Ricci scalar are calculated as
\begin{align}
&K = \frac{(r-\chi) \left(3 r^2+9 r \chi+4 \chi^2\right)+l^2 (2 r+ \chi)\kappa}{l^2 (r+\chi)^2\sqrt{f_n(r)} }, \label{eq:TNPQ0K}\\
&R_3=-\frac{2 \chi^2 \left(\kappa  l^2+r^2+2 r \chi-3 \chi^2\right) \lambda '(\theta )^2 +2 l^2 (r+\chi)^2 Y(\theta ) Y''(\theta ) }{l^2 (r-\chi) (r+\chi)^3 Y(\theta )^2}. \label{eq:TNPQ0R3}
\end{align}
From these we obtain 
\begin{align}
&I_{\textrm{E}} =\frac{\beta \omega}{8 \pi G} \left[3 M_n -2 \kappa  R -\frac{2 \left(R^3-4 R \chi^2-\chi^3\right)}{l^2} + \mathcal{O}\left(\frac{1}{R} \right)\right] +I_{ct} , \\
&I_{\textrm{ct}} = \frac{\beta  \omega}{8 \pi  G l} \sqrt{f_n(R)} \left[2 (R^2 - \chi ^2)+ \kappa  l^2  - \frac{l^2 \chi ^2 f_n(R)}{\left(R^2-\chi ^2\right)} \right],  \label{eq:IctTN}\\
&I_{\textrm{ren.}} = \lim_{R \rightarrow \infty} I_{\textrm{E}} = \frac{\beta  \omega}{8 \pi G} \left(M_n + \frac{2 \chi ^3}{l^2} \right)=  \frac{\beta \omega \chi}{8 \pi G} \bigg(\kappa - \frac{2 \chi^2}{l^2} \bigg)
\end{align}
for the  on-shell Euclidean action (\ref{eq:Iepq0}), 
where  $\omega = \int_0^{2\pi} d\phi \int^{\theta_{max}(\phi)}_{\theta_{min}(\phi)} d \theta Y(\theta)$ is the volume element of the transverse space\footnote{For $\kappa=1$, $\theta_{max}(\phi)=\pi$ and $\theta_{min}(\phi)=0$; for other values of $\kappa$ these quantities depend on the identifications made
in the transverse space \cite{Smith:1997wx}.}, and $\beta$ is the
imaginary time periodicity, which becomes the inverse of temperature.  For non-compact transverse spaces, we shall rescale our thermodynamic quantities by $\omega$.

The free energy is then given by definition
\begin{equation}
F = - \frac{1}{\beta} \log Z \sim \frac{I_{\textrm{ren.}}}{\beta}
\end{equation}
where $Z$ is a partition function and we used a saddle point approximation in the second expression. The thermodynamic entropy and energy are obtained from
\begin{align}\label{ent-eng-free}
S = \bigg(\beta \frac{\partial}{\partial \beta} - 1 \bigg) I_{\textrm{ren.}} , \qquad  E = \partial_{\beta} I_{\textrm{ren.}}
\end{align}
yielding 
\begin{align}
S = \frac{\omega \chi^2}{G} \bigg(\kappa - \frac{6 \chi^2}{l^2} \bigg), \qquad E  = \frac{\chi \omega}{4 \pi G} \bigg(\kappa - \frac{4 \chi^2}{l^2} \bigg) = \frac{\omega}{4 \pi G} M_n.
\end{align}
These quantities satisfy the first law and free energy
\begin{align}
dE = T_n dS, \qquad F = E - T_n S
\end{align}
 which is only valid for $\kappa=1$ case since there is no hyperbolic TN-AdS. For $\kappa=0$, the temperature is zero, we cannot determine the entropy using \eqref{ent-eng-free}.  However since the Misner string is not present, we expect that the entropy satisfies the Bekenstein-Hawking entropy and so becomes the area of the TN-AdS, which is zero.

\subsection{Uncharged TB-AdS}

For TB-AdS, the horizon is   at $r=r_b \neq \chi$ and the mass parameter is  
\begin{align}\label{Mbolt}
M_b = \frac{-6 \chi ^2 r_b^2+r_b^4-3 \chi ^4 + \kappa  \left(r_b^2+\chi ^2\right)l^2}{2 l^2 r_b} 
\end{align}
from \eqref{eucf}.
The extrinsic curvature and the three dimensional Ricci scalar are  
\begin{align}
&K = \frac{(-3 r^2+ \chi ^2) M_b +2 \kappa  r^3 +(3 r^5-14 r^3 \chi ^2+3 r \chi ^4)l^{-2}}{\left(r^2-\chi ^2 \right)^2 \sqrt{f_\textrm{E}(r)} },\\
&R_3=\frac{-2 \chi ^2  \left(-2 M_b r+\kappa  \left(r^2+\chi ^2\right) +(r^4-6 r^2 \chi ^2-3 \chi ^4)l^{-2}\right)\lambda '(\theta )^2}{\left(r^2-\chi ^2\right)^3 Y(\theta )^2}- \frac{2 Y''(\theta )}{\left(r^2-\chi ^2\right) Y(\theta )}.
\end{align}
yielding
\begin{align}
&I_{\textrm{E}} =  \frac{\omega \beta}{8 \pi G} \bigg[ 3M_b -\frac{(2 R^3 + r_b^3-3 r_b \chi^2 )}{l^2}  + \bigg(\frac{8\chi^2}{l^2} - 2 \kappa \bigg)R + \mathcal{O}\left(\frac{1}{R} \right) \bigg] +I_{ct} \label{eq224} \\
&I_{\textrm{ct}} =  \frac{\beta  \omega}{8 \pi  G l} \sqrt{f_\textrm{E}(R)} \left[2 (R^2 - \chi ^2)+ \kappa  l^2  - \frac{l^2 \chi ^2 f_\textrm{E}(R)}{\left(R^2-\chi ^2\right)} \right], \label{eq:IctTB}\\
&I_{\textrm{ren.}} = \lim_{R \rightarrow \infty} I_{\textrm{E}}  = \frac{\omega \beta}{8 \pi G} \left[ M_b - \frac{r_b^3 - 3 r_b \chi^2}{l^2}  \right], \label{eq:IrenNC}
\end{align}
for the Euclidean action.  

\begin{figure}[t!]
\center{\includegraphics[scale=0.5]{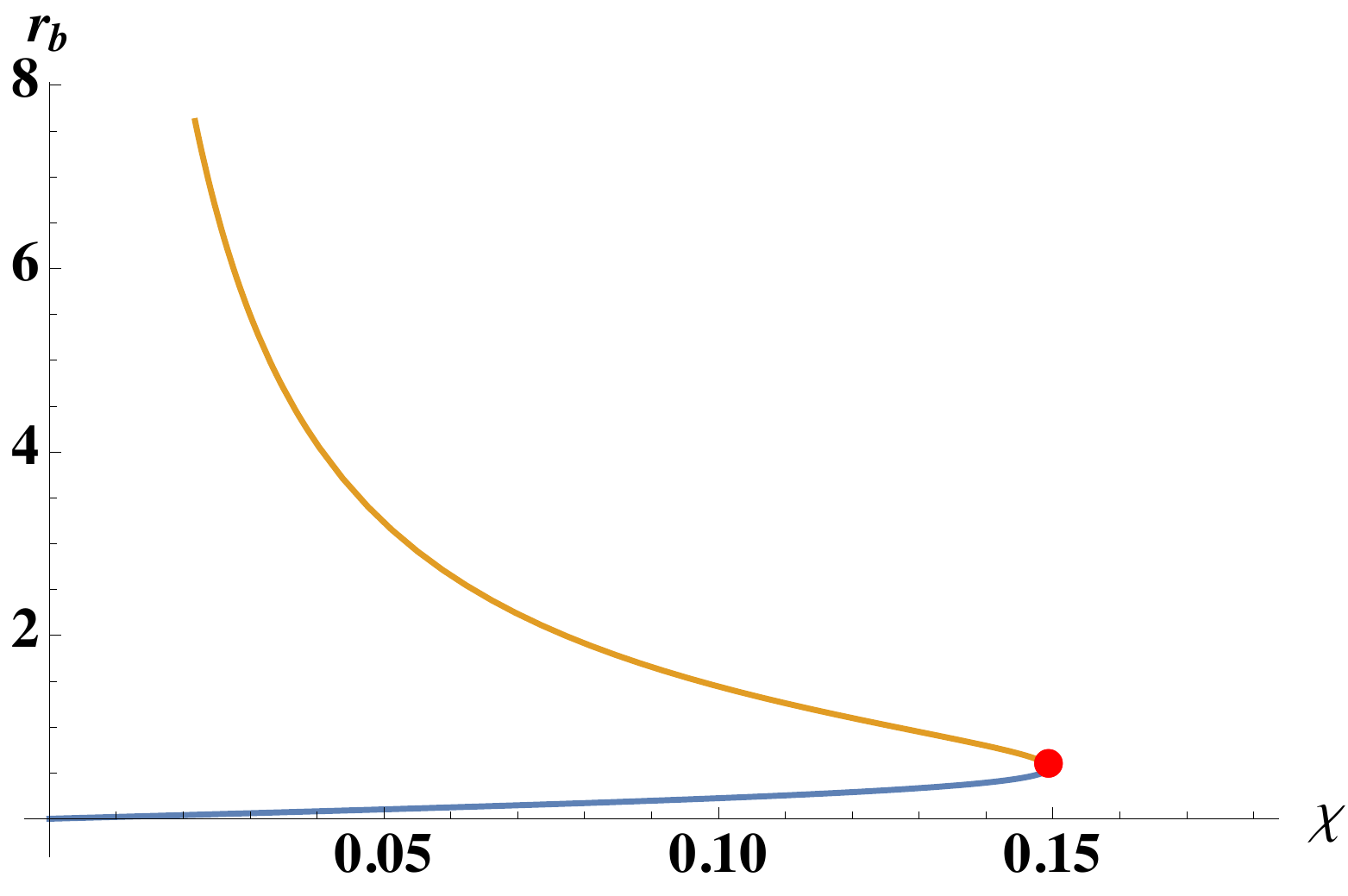}}
\caption{The horizon radius $r_b$ (\ref{eq:rbpq0}) vs NUT charge for uncharged TB-AdS for $\kappa=1$ and $G=l=1$.}
\label{fig:EPQ0}
\end{figure}

The
 Hawking temperature is 
\begin{equation}
T = \frac{1}{4 \pi} f'_{\textrm{E}}(r_b) = \frac{M_b \left(r_b^2+\chi ^2\right) + (9 \chi ^4 r_b-2 \chi ^2 r_b^3+r_b^5)l^{-2}- 2 \kappa  r_b \chi ^2 }{2 \pi  \left(\chi ^2-r_b^2\right)^2}
\label{eq:tbpq0T}
\end{equation}
where  for $\kappa = 1$  we must also have
\begin{equation}
T= \frac{1}{4 \pi} f'_{\textrm{E}}(\chi) = \frac{1}{8 \pi \chi}
\label{eq:tbpq0Tk1}
\end{equation}
to remove the   Misner string singularity. This constrains the thermodynamic phase space. Inserting \eqref{Mbolt} into   (\ref{eq:tbpq0T}) and equating the result to (\ref{eq:tbpq0Tk1}), the horizon radius  is
\begin{equation}
r_b = \frac{1}{12 \chi} \bigg(l^2 \pm \sqrt{l^4 + 144 \chi^4 -48 l^2 \chi^2} \bigg)
\label{eq:rbpq0}
\end{equation}
which we depict in Fig.\ref{fig:EPQ0}  in units with   $G=1$ and $l=1$. 
The blue/orange solid lines correspond to the negative/positive roots in \eqref{eq:rbpq0}. There is a maximum value of the NUT charge at 
\begin{equation}
\chi = \chi_{\textrm{max}} = \frac{l}{2} \sqrt{\frac{1}{3} (2-\sqrt{3})}
\end{equation}
which we denote as a red dot in Fig.\ref{fig:EPQ0}. Thus the NUT charge lies in the range $ 0 \leq \chi \leq \chi_{\textrm{max}}$ and has two branches that show increasing (the lower blue curve) and decreasing (the upper orange curve) behaviour  of the radius as the NUT charge increases.  These two branches have different characteristics in large $l$ limit as follows
\begin{equation}
r_{b-} \sim 2 \chi , \qquad  \qquad r_{b+} \sim \frac{l^2}{6 \chi}.
\end{equation}
The lower branch is smoothly connected TB in flat spacetime when $l$ is taken to infinity, whereas the upper branch does not smoothly go to TB in flat spacetime since $r_b$ diverges in the $l \rightarrow \infty$ limit. Hence the upper branch only exists in AdS.
\\

\begin{figure}[t!]
\center{\includegraphics[scale=0.5]{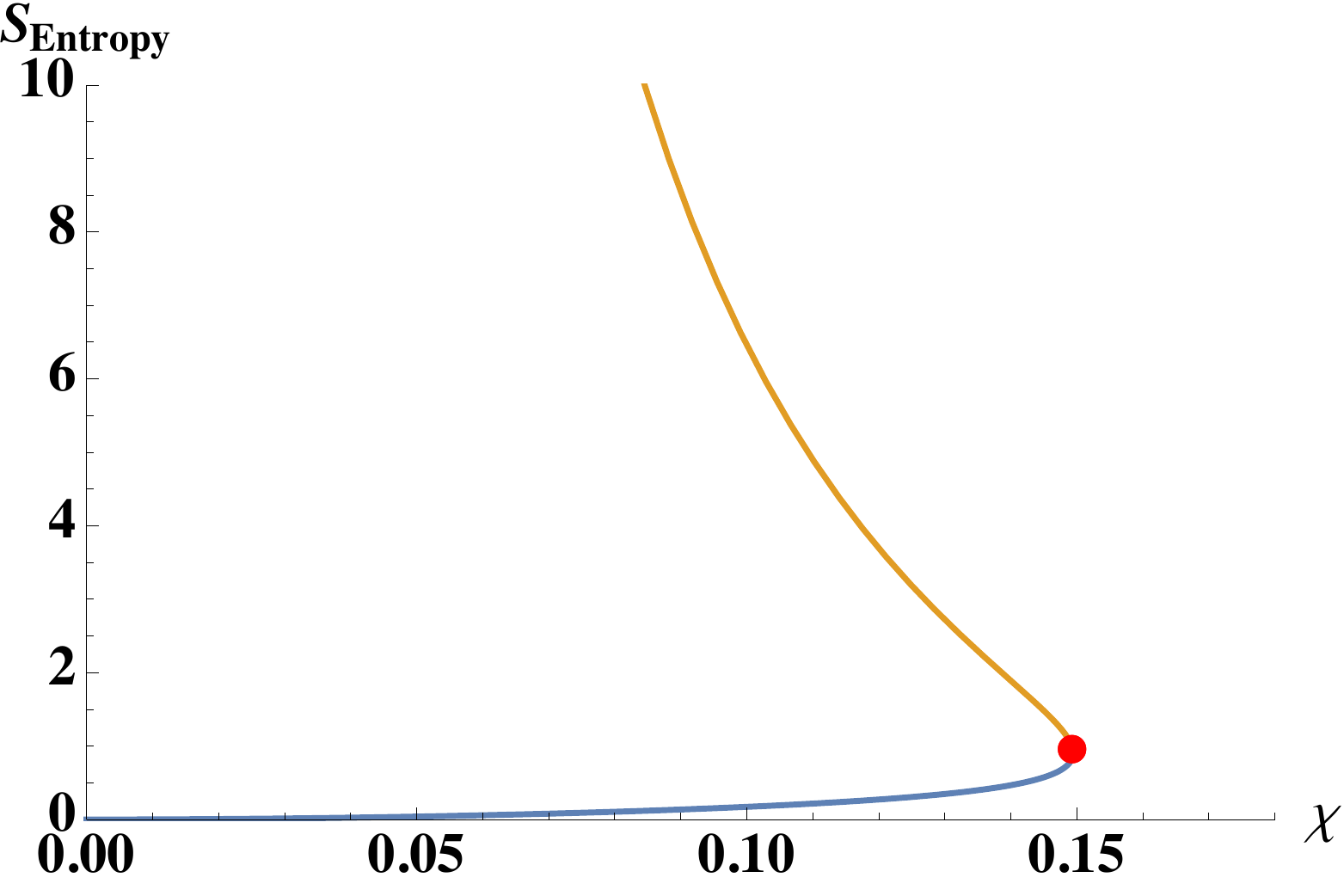}}
\caption{Entropy vs NUT charge $\chi$ for uncharged TB-AdS for $\kappa=1$ and $G=l=1$}
\label{fig:SPQ0}
\end{figure}

From the renormalized   action \eqref{eq:IrenNC}, we calculate thermodynamic energy, entropy, and specific heat   using the definitions
\begin{align}
E = \partial_{\beta} I_{\textrm{ren.}}, \qquad S = \bigg(\beta \frac{\partial}{\partial \beta} - 1 \bigg) I_{\textrm{ren.}}, \qquad C = - \beta \frac{\partial S}{\partial \beta}.
\label{eq:ESC}
\end{align}
For $\kappa=1$, the energy and entropy become 
\begin{align}
&E_{(\kappa=1)}= \frac{\omega}{8 \pi G l^2 r_b} \bigg(r_b^4 - 6 r_b^2 \chi^2- 3 \chi^4 + l^2 (r_b^2+\chi^2) \bigg) = \frac{\omega}{4 \pi G} M_b, \\
&S_{(\kappa=1)} = \frac{\omega}{4 G r_b}  \left(r_b^3-r_b \chi ^2 + 4 \kappa  \chi ^3-\frac{12 \chi ^3 \left(r_b^2+\chi ^2\right)}{l^2}\right) = \frac{\omega \chi}{G} \bigg(M_b+ \frac{r_b^3 - 3 r_b \chi^2}{l^2} \bigg)
\label{eq:Spq0k1}
\end{align}
where the entropy is not the area of the TB-AdS horizon due to the contribution of Misner string. Inserting (\ref{eq:rbpq0}) into (\ref{eq:Spq0k1}), we plot the entropy as a function of $\chi$ in Fig.\ref{fig:SPQ0}. We see that the entropy increases for the small TB-AdS (blue solid line) solution and decreases for the large one (orange solid line) as the NUT charge $\chi$ increases to its maximum value. 

\begin{figure}[b!]
\vspace{0.5cm}
\center{\includegraphics[scale=0.5]{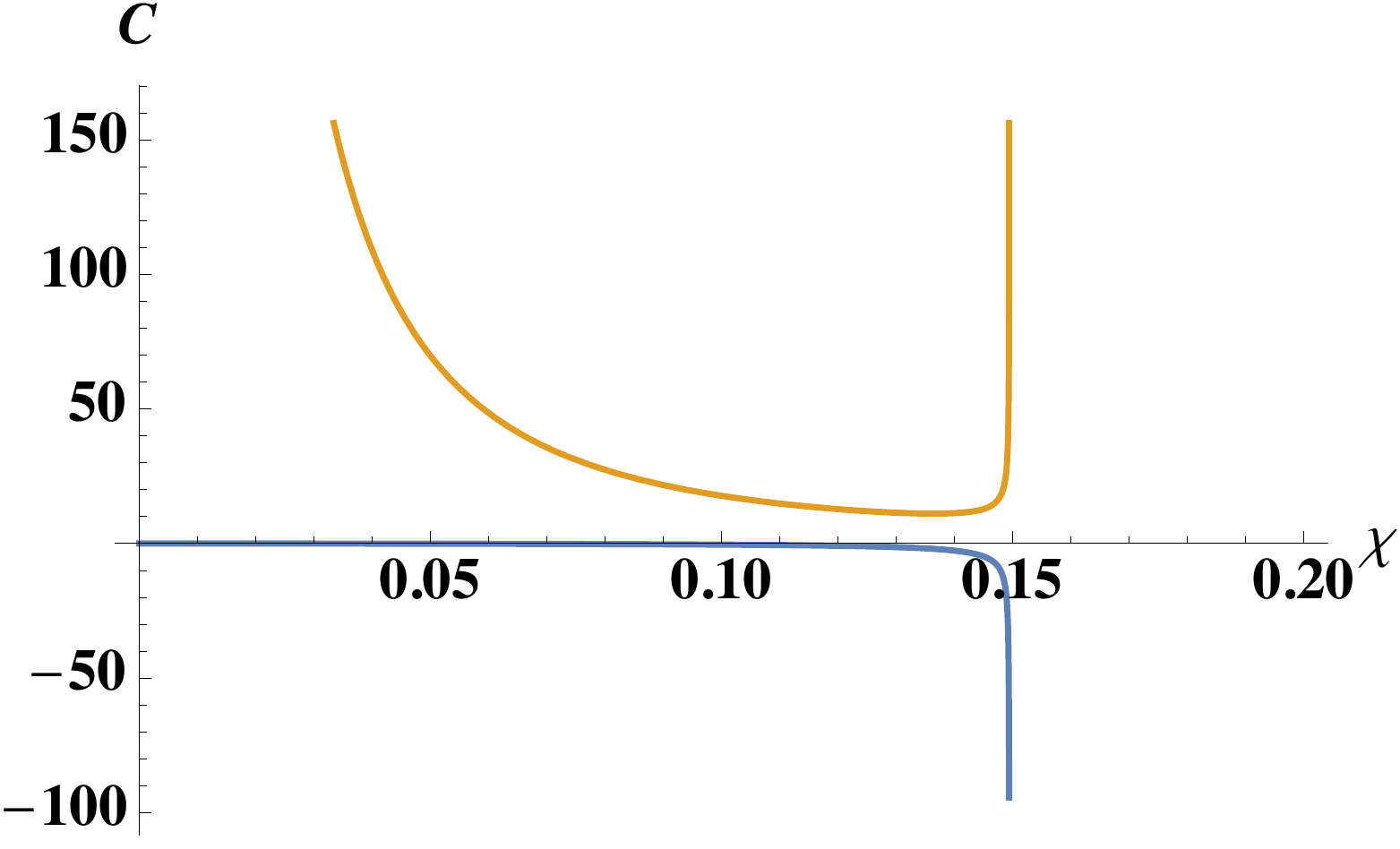}}
\caption{Heat Capacity (\ref{eq:Cpq0k1}) vs NUT charge for uncharged TB-AdS for $\kappa=1$ and  {$G=l=1$}}
\label{fig:CPQ0}
\end{figure}

For $\kappa \neq 1$, the energy and entropy are  
\begin{align}
&E_{(\kappa \neq1)}= \frac{\omega}{8 \pi G}\bigg(\kappa + \frac{r_b^3 - 3 r_b \chi^2}{l^2}  \bigg) ,\\
&S_{(\kappa \neq1)}=\frac{\omega}{4 G} (r_b^2 - \chi^2)
\end{align}
and the entropy agrees with the Bekenstein-Hawking formula. These quantities for all $\kappa$ satisfy the first law and free energy
\begin{equation}
dE = T d S, \qquad F= E - T S.
\end{equation}
Finally, the heat capacity 
\begin{align}
C = T \frac{\partial S}{\partial T} = T \bigg(\frac{\partial T}{\partial M} \frac{\partial M}{\partial r_b} \frac{\partial r_b}{\partial S} \bigg)^{-1}
\label{eq:C}
\end{align}
is
\begin{align}
&C_{(\kappa=1)} = \frac{\omega}{2 G} \frac{r_b(12 \chi^3 - l^2 r_b)}{(-12 r_b \chi + l^2)}, \label{eq:Cpq0k1}\\
&C_{(\kappa \neq 1)} = \frac{\omega r_b^2}{2 G} \bigg(\frac{3 r_b^2-3 \chi^2 + \kappa  l^2 }{3 r_b^2+3 \chi^2 -\kappa  l^2} \bigg)
\end{align}
for the different values of $\kappa$.  We plot the  $\kappa=1$ case as a function of   $\chi$  (using (\ref{eq:rbpq0}))   in Fig.\ref{fig:CPQ0}. Similar to  small and large AdS black holes, the small (blue solid line) and large (orange solid line) TB-AdS solutions exhibit negative and positive heat capacity respectively. 

\newpage
\section{Thermodynamics  of Dyonic TN/TB-AdS}

To obtain dyonic Taub-NUT-AdS solutions, we add a gauge field to the action \eqref{fullact}
\begin{equation}
I = \frac{1}{16 \pi G_4} \int d^4 x \sqrt{-g} \bigg(R + \frac{6}{l^2} - F^{\mu \nu} F_{\mu \nu} \bigg) + \frac{1}{8 \pi G_4} \int d^3 x \sqrt{-h} \bigg(K- \frac{2}{l} - \frac{l}{2} R_3 \bigg)
\end{equation}
which leads to the equations of motion  
\begin{align}
&R_{\mu \nu} - \frac{1}{2} g_{\mu \nu} R = \frac{3}{l^2} g_{\mu \nu} + 2 \bigg(F_{\mu}^{\; \sigma} F_{\nu \sigma} - \frac{1}{4} g_{\mu \nu} F_{\rho \sigma} F^{\rho \sigma} \bigg), \\
&D_{\mu} F^{\mu \nu} = 0. \label{eq:DFeom}
\end{align}
The  solution for the metric and gauge field are  
\begin{align}
&ds^2 = -f(r)(d t + 2 s \lambda(\theta) d \phi)^2 + \frac{dr^2}{f(r)} + (r^2 + s^2) (d \theta^2 + Y(\theta)^2 d \phi^2), \\
&A = A_{\mu} dx^{\mu} = \frac{1}{2 s} h(r) d t + h(r) \lambda(\theta) d\phi \label{eq:A}
\end{align}
where
\begin{align}
&f(r) = \frac{l^{-2}(r^2 + s^2)^2 + (\kappa + 4 l^{-2} s^2)(r^2 - s^2) - 2 M r + P^2 + Q^2}{r^2 + s^2},\\
&h(r) = - \frac{2 Q s r + P (r^2 - s^2) }{r^2 + s^2}. 
\label{eq:hofA}
\end{align}
From the  field strength of the gauge field (\ref{eq:A}) we obtain the electric and magnetic charges
\begin{align}
&Q_e [\xi_t] \equiv \frac{1}{\omega} \int_{\partial {\Sigma}_{\kappa}} *F = \lim_{r \rightarrow \infty} \frac{Q \left(r^2-s^2\right)-2 P r s}{r^2+s^2} = Q, \\
&Q_m[\xi_{\phi}] \equiv \frac{1}{\omega} \int_{\partial {\Sigma}_{\kappa}}  F =\lim_{r \rightarrow \infty}  \frac{P \left(r^2-s^2\right)+2 Q r s}{r^2+s^2} = P
\label{eq:emcharge}
\end{align}
where $\Sigma_{\kappa}$ is the spacelike hypersurface for each $\kappa$, $\partial {\Sigma}_{\kappa}$ is the boundary of $\Sigma_{\kappa}$ at $r=$constant, and the conserved charge is taken at $r=\infty$.

Once again, performing the   Wick-rotation $t \rightarrow - i \tau$ and $s \rightarrow - i \chi$, we find 
\begin{align}
&I_{\textrm{ren.}} = - \frac{1}{16 \pi G_4} \int d^4 x \sqrt{g} \bigg(R + \frac{6}{l^2} - F^{\mu \nu} F_{\mu \nu} \bigg) - \frac{1}{8 \pi G_4} \int d^3 x \sqrt{h} \bigg( K - \frac{2}{l} - \frac{l}{2} R_3 \bigg)  
\end{align}
for the Euclidean action.  The metric and gauge field become
\begin{align}
&ds_{\textrm{E}}^2 = f_E(r)(d \tau + 2 \chi \lambda(\theta) d \phi)^2 + \frac{dr^2}{f_E(r)} + (r^2 - \chi^2) (d \theta^2 + Y(\theta)^2 d \phi^2) \label{eq:elds} \\
&A_{\textrm{E}}  =  \frac{1}{2 \chi} h_E(r) d \tau + h_E(r) \lambda(\theta) d\phi, \label{eq:elA}
\end{align}
where
\begin{align}
&f_{\textrm{E}} = \frac{l^{-2}(r^2 - \chi^2)^2 + (\kappa - 4 l^{-2} \chi^2)(r^2 + \chi^2) - 2 M r + P^2 + Q^2}{r^2 - \chi^2}, \label{eq:TNgE} \\
&h_{\textrm{E}}  = \frac{2 i Q \chi r - P (r^2 + \chi^2) }{r^2 - \chi^2}. \label{eq:TNhE}
\end{align}
As with the uncharged case, this Euclidean metric describes two distinct geometries, a dyonic-TN-AdS and a dyonic-TB-AdS.

\subsection{Dyonic TN-AdS solution and thermodynamics}

The dyonic-TN-AdS solution is obtained  when
  $f_{\textrm{E}}(r)$ has a root at  $r=r_+ = \chi$.  This implies
\begin{equation}\label{Mnchg}
M=M_{n} = \frac{1}{2 \chi} \bigg(2 \chi^2 (\kappa - \frac{4 \chi^2}{l^2}) + P^2 + Q^2 \bigg).
\end{equation}
and the  Euclidean metric function   is then given by
\begin{equation}
f_n(r) = \frac{l^{-2}(r-\chi)^2 (r+3\chi)\chi + \kappa  (r-\chi) \chi  - (P^2 + Q^2)}{\chi(r+\chi)} \; . \label{eq:PQfn}
\end{equation}
It is clear that if $f_n(r=\chi) = 0$, then
\begin{equation}
P^2 + Q^2 = 0 \label{eq:TNPQ2}
\end{equation}
We can satisfy this relation non-trivially if we   take $P=iQ$ so as to make the gauge field vanish at the horizon, namely $h_E(r_+)=h(\chi)=0$. This implies
\begin{equation}
h_{\textrm{E}}(r)=- \frac{ i Q(r-\chi)}{r+\chi}.
\end{equation}
Note that if   (\ref{eq:TNPQ2}) holds then the metric function becomes the same as the uncharged case, and so only $\kappa=0$ and $1$ are allowed for dyonic TN-AdS.   Imposing these conditions $r_+= \chi$ and $P=iQ$, the mass parameter becomes
\begin{equation}
M_n = \chi \kappa - \frac{4 \chi^3}{l^2}.
\end{equation}
Since $K$ and $R_3$ are still given by (\ref{eq:TNPQ0K}) and (\ref{eq:TNPQ0R3}) respectively, the on-shell Euclidean action is now
\begin{align}
I_{\textrm{E}} &= \frac{\beta \omega}{8 \pi G} \left[ 3 M_n -2 \kappa  R -\frac{2 \left(R^3-4 R \chi^2-\chi^3\right)}{l^2}-\frac{Q^2}{2 \chi} + \mathcal{O}\left(\frac{1}{R} \right) \right] + I_{ct}, \\
I_{\textrm{ren.}} & = \lim_{R \rightarrow \infty} (I_{\textrm{E}} )  = \frac{\beta  \omega}{8 \pi G}  \left(M_n +\frac{2 \chi ^3}{l^2}-\frac{Q^2}{2 \chi }\right)=  \frac{\beta \omega \chi}{8 \pi G} \bigg(\kappa - \frac{2 \chi^2}{l^2} -\frac{Q^2}{2 \chi^2} \bigg) \label{eq:IrenTN}
\end{align}
where $\beta$  is the inverse of the Hawking temperature.    

The thermodynamic energy, entropy, and specific heat are, for
 $\kappa=1$,  
\begin{align}
&S = \frac{\omega \chi^2}{G} \bigg(1 - \frac{6 \chi^2}{l^2} + \frac{Q^2}{2 \chi^2} \bigg) = \frac{\omega  \chi ^2}{G}  \left(1+ 2 g^2-\frac{6 \chi ^2}{l^2}\right), \label{eq:STN}\\
&E = \frac{\omega}{8 \pi G} \bigg(M_n + \chi \kappa - \frac{4 \chi^3}{l^2} + \frac{Q^2}{\chi} \bigg) = \frac{\omega \chi}{4 \pi G}  \left(1+ 2 g^2-\frac{4 \chi ^2}{l^2}\right), \label{eq:ETN} \\
&\Phi =  \frac{i \omega}{4 \pi G} \frac{1}{2 \chi} \bigg(h_{\textrm{E}}(\infty) -h_{\textrm{E}}(\chi)  \bigg) = - \frac{\omega }{4 \pi G} g. \label{eq:PhiTN}
\end{align} 
where we rescaled $Q = - 2 g \chi$ (and so $P = - 2 i g \chi$). The Hawking temperature is easily calculated to be        
\begin{equation}
T = \frac{1}{4 \pi} f'_n (\chi) = \frac{\kappa}{8 \pi \chi}
\end{equation}
and we can verify that the first law and free energy are satisfied
\begin{equation}
dE = TdS + \Phi dQ, \qquad F = E - TS - \Phi Q.
\end{equation}
where  $\chi$ and $g$ are the independent variables. 

For $\kappa=0$ the temperature vanishes and we have an extremal solution. 
From the gauge potential, we expect the energy to be  
\begin{equation}
E = \frac{\omega}{4 \pi G} \left(\frac{Q^2}{4 \chi }-\frac{\chi ^3}{l^2}\right), \qquad \Phi = \frac{Q \omega}{8 \pi G \chi} \label{eq:ETNext}
\end{equation}
so as to satisfy the following relations
\begin{equation}
dE = \Phi dQ, \qquad F = E  - \Phi Q.
\end{equation}
Since the temperature is zero, we cannot determine the entropy using \eqref{ent-eng-free}.  However the horizon area is zero and the Misner string regularity condition does not apply, so we expect that the entropy is  zero.

\subsection{Dyonic TB-AdS}

If we assume that the metric (\ref{eq:TNgE}) is not factored by ($r-\chi$), then it has a two dimensional fixed point set and is called TB-AdS. The metric and gauge field are now
\begin{align*}
&f_E = \frac{l^{-2}(r^2 - \chi^2)^2 + (\kappa - 4 l^{-2} \chi^2)(r^2 + \chi^2) - 2 M r + P^2 + Q^2}{r^2 - \chi^2}, \\
&h_{\textrm{E}}  = \frac{2 i Q \chi r - P (r^2 + \chi^2) }{r^2 - \chi^2}. 
\end{align*}
and the mass parameter is  
\begin{equation}
M=M_b = \frac{l^2 \left(P^2+Q^2+\kappa  \left(r_b^2+\chi ^2\right)\right)+r_b^4-6 r_b^2 \chi ^2-3 \chi ^4}{2 l^2 r_b} \label{eq:Mb}
\end{equation}
where we denote the radius of the dyonic-TB-AdS as $r_b$. Imposing the regularity condition for the norm of a gauge field to vanish at the horizon of TB-AdS, we get	
	\begin{equation}
	g^{a b} A_{E a} A_{E b} \big|_{r=r_b} = A_E^2(r_b) = 0 \; \;  \rightarrow \; \; P = \frac{2 i Q \chi  r_b}{r_b^2+\chi ^2}
	\label{eq:regcd1}
	\end{equation}
and so $Q$ and $P$ are related. 

We next calculate all thermodynamic quantities, subsequently apply the regularity condition, and then show the first law and free energy are satisfied. 
To   calculate the electric and magnetic potentials, we consider two definitions of the field potential as follows
\begin{equation}
\Phi_{E}^{(1)} = \frac{\partial E}{\partial Q} \bigg|_{r_b,P} - T \frac{\partial S}{\partial Q}\bigg|_{r_b,P}, \qquad \Phi_M^{(1)} = \frac{\partial E}{\partial P} \bigg|_{r_b,Q} - T \frac{\partial S}{\partial Q}\bigg|_{r_b,Q}
\label{eq:emp1}
\end{equation}
 where $\Phi_{E}^{(1)} $ and $\Phi_{M}^{(1)}$ are conjugate variables for each charges $Q$ and $P$ respectively via the first law of thermodynamics
\begin{equation}
dE = TdS + \Phi_{E} dQ + \Phi_{M} dP
\end{equation}
with details about how to compute them given in appendix \ref{sec:appGP}. Alternatively, we can take
\begin{align}
\Phi_E^{(2)} = \frac{i \omega}{4 \pi G} ( A_{\tau}(\infty)-A_{\tau}(r_b)), \qquad \Phi_M^{(2)} = \frac{\omega}{4 \pi G} (\Phi_M(\infty) - \Phi_M(r_b))  \label{eq:emp2}
\end{align} 
 which are typically chosen for gauge potentials, and where  
\begin{equation}\label{Phi2M}
\Phi_M (r) = \int^r dr' B(r'), \qquad B(r) = \frac{1}{\sqrt{g}} \epsilon^{t r \theta \phi} F_{\theta \phi}\; .
\end{equation}
Then we can compute energy from the equation
\begin{equation}
E  - \Phi_ {\textrm{E}} Q - \Phi_{\textrm{M}} P = \partial_{\beta} I_{\textrm{ren.}}
\label{eq:defE}
\end{equation}
 for both cases.

Since we have a dyonic solution we naturally expect both types of charge to appear in the first law, and usually they yield the same
result  {for an electric potential, as shown in appendix \ref{sec:appRNbh}}.  Here, however, this turns out not to be the case:  upon computing the potentials respectively using \eqref{eq:emp1} and \eqref{eq:emp2} for each case and then imposing the regularity condition (\ref{eq:regcd1}),  we find that thermodynamic quantities such as energy, temperature, entropy, and electric potentials  $\Phi_E^{(1)}$ and $\Phi_E^{(2)}$ become equivalent, whereas the magnetic potentials $\Phi_M^{(1)}$ and $\Phi_M^{(2)}$ remain inequivalent. This is because the regularity condition is equivalent to setting the magnetic potential $\Phi_M^{(1)}=0$,  a condition incompatible with  $\Phi_M^{(2)}$ in \eqref{eq:emp2}. Thus   the magnetic potential $\Phi_M^{(2)}$ is not valid once the regularity condition (\ref{eq:regcd1}) is imposed. Then under this regularity condition (\ref{eq:regcd1}) we expect the first law and free energy to satisfy as follows
\begin{align}
dE = TdS + \Phi_{E} dQ, \qquad F = E - TS - \Phi_E Q
\end{align}
and we shall illustrate this in specific cases in what follows. In general only $\Phi_M^{(1)}$ can satisfy the first law, as should be clear from its definition in 
\eqref{eq:emp1}.

\subsubsection{Thermodynamics for $\kappa \neq 1$}

The extrinsic curvature and three dimensional Ricci scalar are  
\begin{align}
&K = \frac{-3 M_b r^2+M_b \chi ^2+2 \kappa  r^3 + r \left(P^2+Q^2\right)+ (3 r^5-14 r^3 \chi ^2+3 r \chi ^4)l^{-2}}{\left(r^2-\chi ^2\right)^2 \sqrt{f_{\textrm{E}}(r)}},\\
&R_3 = \frac{-2 \chi ^2  \left(-2 M_b r+P^2+Q^2+\kappa  \left(r^2+\chi ^2\right) +(r^4-6 r^2 \chi ^2-3 \chi ^4)l^{-2}\right)\lambda '(\theta )^2}{\left(r^2-\chi ^2\right)^3 Y(\theta )^2}- \frac{2 Y''(\theta )}{\left(r^2-\chi ^2\right) Y(\theta )} 
\end{align}
and from this we obtain
\begin{align} 
I_{\textrm{E}} & = \frac{\beta  \omega}{8 \pi G}  \left[3 M_b -2 \kappa  R -\frac{(2 R^3-8 R \chi ^2+r_b^3-3 r_b \chi ^2)}{l^2}-\frac{r_b \left(\left(Q^2-P^2\right) \left(r_b^2+\chi ^2\right)+4 i P Q r_b \chi \right)}{\left(r_b^2-\chi ^2\right)^2} \right.  \nonumber\\
&\qquad  \left.+ \mathcal{O}\left(\frac{1}{R} \right) \right] +I_{\textrm{ct}} \\  
I_{\textrm{ren}} &= \lim_{R \rightarrow \infty} (I_{\textrm{E}})  = \frac{\beta  \omega}{8 \pi G}  \left(M_b -\frac{r_b^3-3 r_b \chi^2}{l^2}+\frac{\left(P^2-Q^2\right) \left(r_b^2+\chi^2\right)r_b -4 i P Q r_b^2 \chi }{\left(r_b^2-\chi^2\right)^2}\right).  
\label{eq:IETB}
\end{align}
for  the Euclidean action.  Note from \eqref{eq:regcd1} that   $P \propto - i 2 Q \chi$, ensuring  the Euclidean action   (\ref{eq:IETB}) is real. 

The Hawking temperature and entropy are  
\begin{align}
&T = \frac{1}{4 \pi} f'_E(r_b) \nonumber\\
&=\frac{r_b  \chi ^2}{\pi  \left(r_b^2-\chi ^2\right)^2}  \left(\frac{M_b \left(r_b^2+\chi ^2\right)}{2 r_b  \chi ^2} +\frac{r_b^4-2 r_b^2 \chi ^2+9 \chi ^4}{ 2 l^2 \chi^2} -\frac{(P^2+Q^2)}{ 2\chi^2} -\kappa \right) \nonumber  \\
&= \frac{l^2 (\kappa  r_b-M_b)+2 r_b \left(r_b^2-3 \chi^2 \right)}{2 \pi l^2 \left(r_b^2-\chi^2\right)}  \label{eq:etnpTBkn2} \\
&S \equiv  \bigg(\beta \frac{\partial}{\partial \beta}- 1 \bigg)I_{\textrm{ren.}} = \frac{\omega  \left(r_b^2-\chi^2\right)}{4 G}, \label{eq:etnpTBkn1}
\end{align}
where 
 \begin{equation}
\beta = \frac{2 \pi  l^2 \left(r_b^2-\chi ^2\right)}{l^2 \left(\kappa  r_b-M_b \right)+2 r_b \left(r_b^2-3 \chi ^2\right)}
\end{equation}
allows us to regard $M_b$  as a function of $\beta$ and the other parameters. The heat capacity for fixed $P$ and $Q$ is 
\begin{equation}
C_{P,Q} = \frac{\omega  r_b^2 \left(r_b^2-\chi ^2\right) \left(3 \left(r_b^2-\chi ^2\right){}^2-l^2 \left(-\kappa  r_b^2+\kappa  \chi ^2+P^2+Q^2\right)\right)}{2 G \left(l^2 \left(\left(P^2+Q^2\right) \left(3 r_b^2-\chi ^2\right)-\kappa  \left(r_b^2-\chi ^2\right){}^2\right)+3 \left(r_b^2-\chi ^2\right){}^2 \left(r_b^2+\chi ^2\right)\right)}
\end{equation}
 from (\ref{eq:C}).
Now we obtain the electric, magnetic potentials and energy 
\begin{align}
&\Phi_{\textrm{E}}^{(1)}= \frac{\omega}{4 \pi G} \frac{Q \left(r_b^2+\chi^2\right) r_b+2 i P r_b^2 \chi}{\left(r_b^2-\chi^2\right)^2}, \qquad \Phi_{\textrm{M}}^{(1)} = \frac{\omega}{4 \pi G } \frac{2 i Q r_b^2 \chi - P \left(r_b^2+\chi^2\right)r_b }{\left(r_b^2-\chi^2\right)^2} \label{eq:EGP1} \\
&E^{(1)} = \frac{r_b \omega}{8 \pi  G} \left(\kappa +\frac{r_b^2-3 \chi^2}{l^2}+\frac{\left(Q^2-P^2\right) \left(r_b^2+\chi^2\right)+4 i P Q r_b \chi}{\left(r_b^2-\chi^2\right)^2}\right), \label{eq:EAem1}
\end{align}
where $\Phi_{\textrm{E}}^{(1)}$ and $\Phi_{\textrm{M}}^{(1)}$ are derived from (\ref{eq:emp1}) and the energy is from (\ref{eq:defE}), and 
\begin{align}
&\Phi_{\textrm{E}}^{(2)} = \frac{\omega}{4 \pi G} \frac{Q r_b +i P \chi}{\left(r_b^2- \chi^2\right)} , \qquad \Phi_{\textrm{M}}^{(2)}  =\frac{\omega}{4 \pi G}\frac{i Q \chi -P r_b}{\left(r_b^2- \chi^2\right)}.  \label{eq:EGP2} \\
&E^{(2)} = \frac{\omega}{8 \pi G}  \left(\kappa  r_b  + \frac{r_b^3-3 r_b \chi^2}{l^2}+\frac{(P^2 r_b -2 i P Q \chi) \left(r_b^2+\chi^2\right)+Q^2 r_b \left(r_b^2-3 \chi^2\right)}{\left(r_b^2-\chi^2\right)^2}\right). \label{eq:EAem2} 
\end{align}
where $\Phi_{\textrm{E}}^{(2)}$ and $\Phi_{\textrm{M}}^{(2)}$ are derived from (\ref{eq:emp2}) and the energy is computed from (\ref{eq:defE}). 
These quantities do not satisfy the  {first law of thermodynamics}, but upon imposing the regularity condition (\ref{eq:regcd1}) these thermodynamic variables (\ref{eq:EGP1})-(\ref{eq:EAem2}) become
\begin{align}
&E^{(1)} = E^{(2)} = E = \frac{\omega}{8 \pi G}  \left(\kappa  r_b + \frac{r_b \left(r_b^2-3 \chi ^2\right)}{l^2}+\frac{Q^2 r_b}{r_b^2+\chi ^2}\right), \label{eq:Ekn1}\\
&\Phi_{\textrm{E}}^{(1)} =\Phi_{\textrm{E}}^{(2)}=\Phi_{\textrm{E}} = \frac{\omega}{4 \pi G}\frac{Q r_b}{\left(r_b^2+\chi ^2\right)} \qquad 
\qquad \Phi_{\textrm{M}}^{(1)}  = 0 \neq \Phi_{\textrm{M}}^{(2)}  = - \frac{\omega}{4 \pi G}\frac{i Q \chi}{\left(r_b^2+\chi ^2\right)}
\end{align}
and we see that the magnetic potentials are inequivalent.
The first law and the free energy become
\begin{align}
&dE = T d S + \Phi_{\textrm{E}} dQ ,\\
&F = E - T S - \Phi_{\textrm{E}} Q
\end{align} 
upon using  $\Phi_{\textrm{M}}^{(1)} = \Phi_{\textrm{M}}  = 0$.

\subsubsection{Thermodynamics for $\kappa = 1$}

For $\kappa = 1$, the Hawking temperature is constrained to be $1/8 \pi \chi$ so that the Misner string singularity is no longer present.
Moreover, the  entropy does not follow the Bekenstein-Hawking formula, but can be obtained from the thermodynamic relation (\ref{eq:ESC}).
We find
\begin{align}
&T = \frac{1}{4 \pi} f'_E(r_b) \\
&= \frac{r_b  \chi ^2}{\pi  \left(r_b^2-\chi ^2\right)^2}  \left(\frac{M_b \left(r_b^2+\chi ^2\right)}{2 r_b  \chi ^2} +\frac{r_b^4-2 r_b^2 \chi ^2+9 \chi ^4}{ 2 l^2 \chi^2} -\frac{(P^2+Q^2)}{ 2\chi^2} -1 \right) = \frac{1}{8 \pi \chi}, \label{eq:TTB} \\
&S \equiv  \bigg(\beta \frac{\partial}{\partial \beta}-1\bigg)I_{\textrm{E}}=\bigg(\chi \frac{\partial}{\partial \chi}- 1 \bigg)I_{\textrm{E}} \label{eq:Sk1L} \\
& = \frac{\omega}{2 G} \left(\frac{l^2 \left(r_b^2+\chi^2\right)-24 r_b \chi^3}{2 l^2}+\frac{4 r_b \chi^2 \left(\left(P^2-Q^2\right) \left(3 r_b^2 \chi+\chi^3\right)-2 i P Q r_b \left(r_b^2+3 \chi^2\right)\right)}{\left(r_b^2-\chi^2\right)^3}\right).\label{eq:etnpTBk1}
\end{align}
without imposing \eqref{eq:regcd1}. We shall first consider the behaviour of these quantities without this constraint, and later see that it is required in order that the  {first law of thermodynamics} hold.

\begin{figure}[t!]
\center{\includegraphics[scale=0.45]{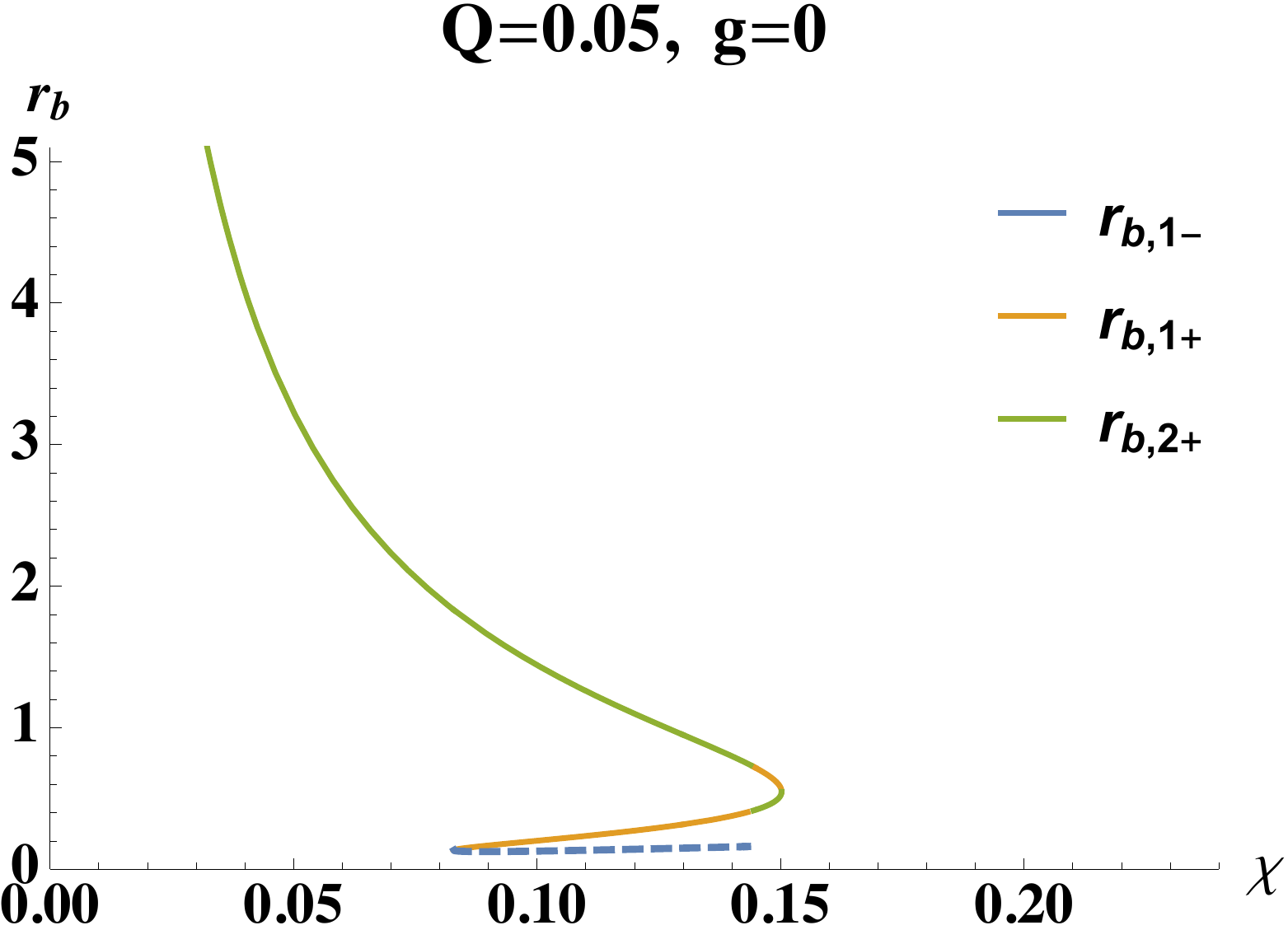} \; \; \includegraphics[scale=0.45]{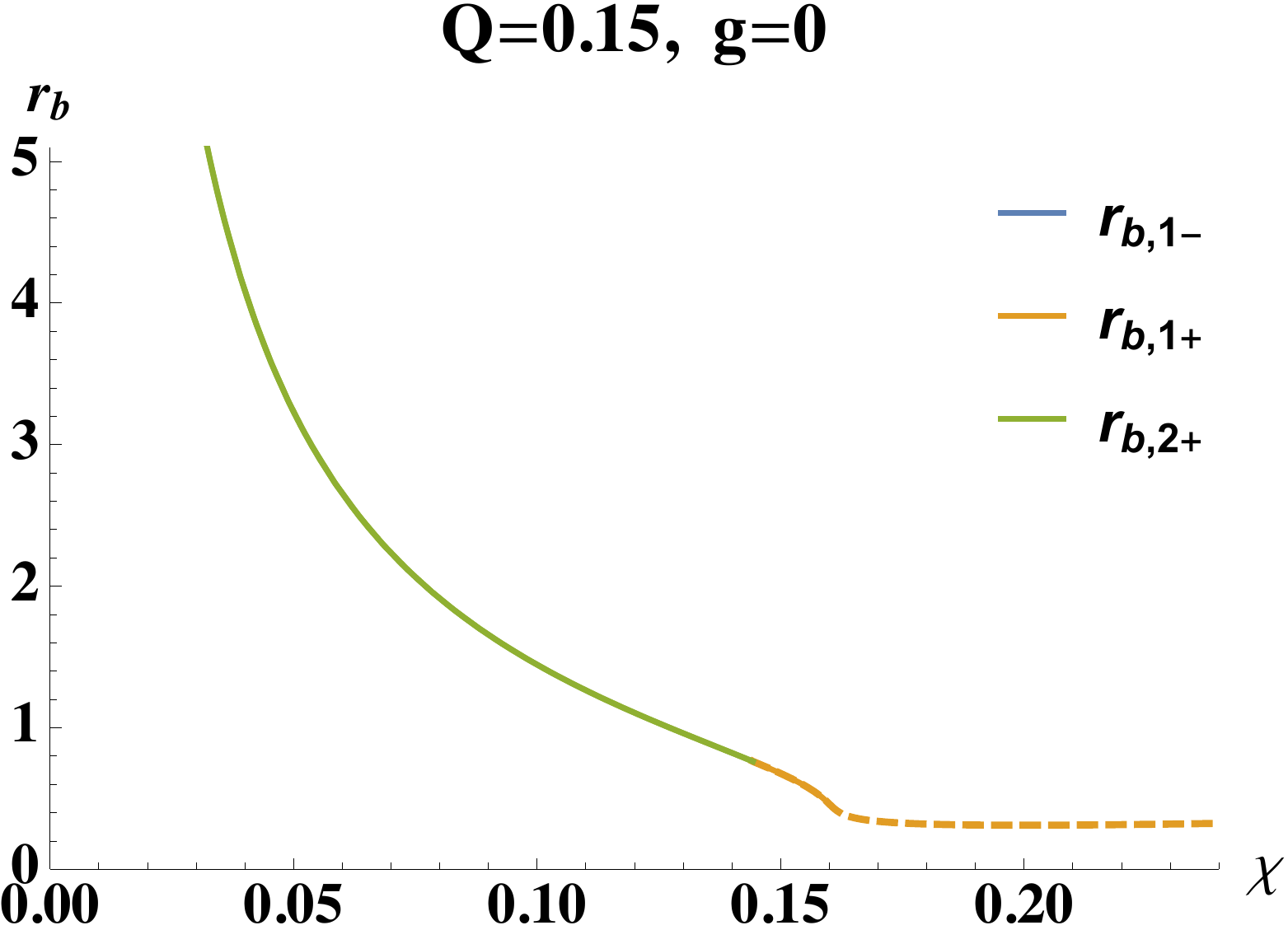} \\
\vspace{0.5cm}
\includegraphics[scale=0.45]{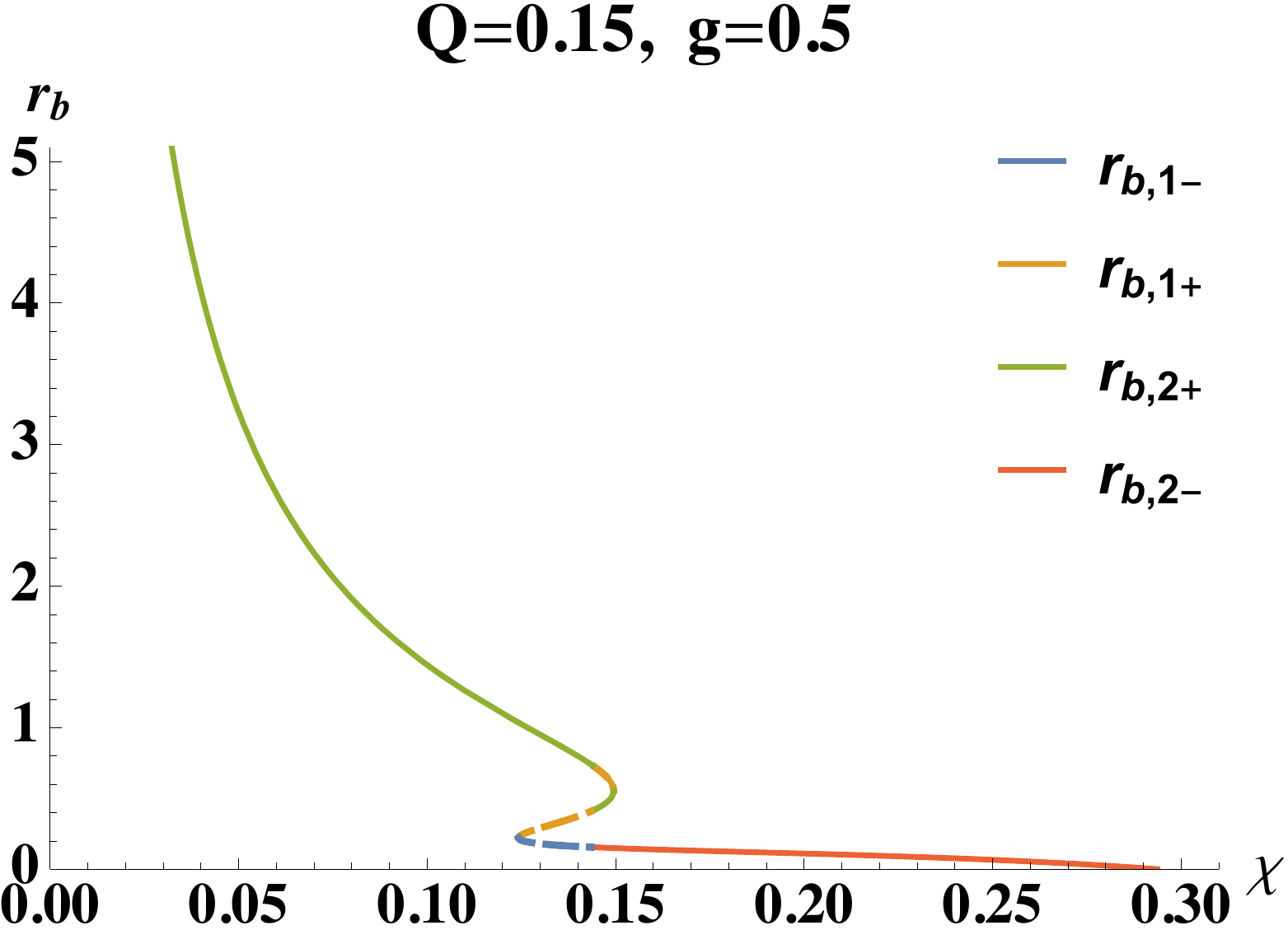} \; \; \includegraphics[scale=0.45]{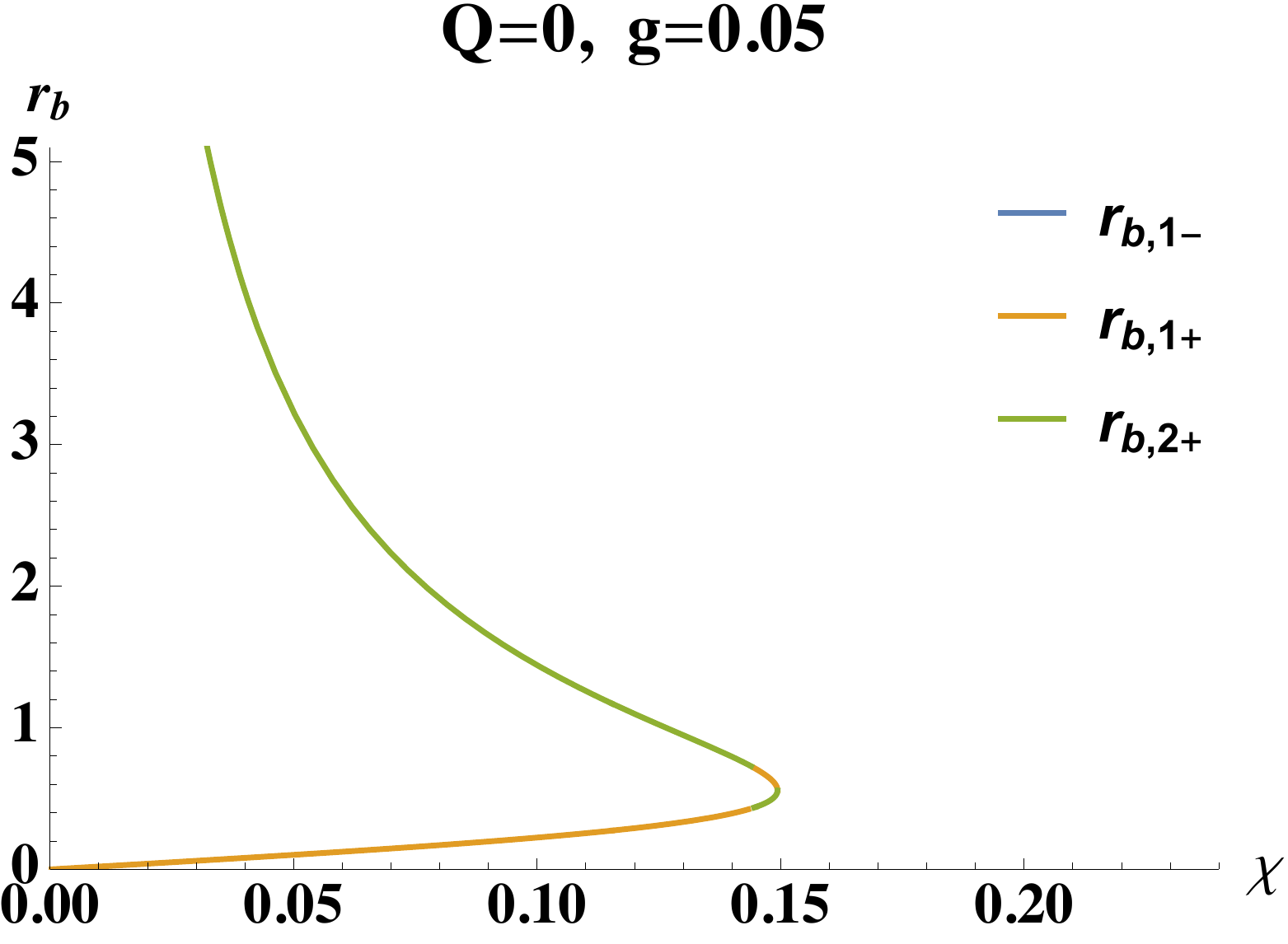}}
\caption{The horizon radius of TB-AdS for $\kappa=1$ vs $\chi$ for fixed values of $Q$ and $g$ with $G=l=1$. The dashed lines correspond to negative entropy, and are not regarded as corresponding to physically admissible solutions.}
\label{fig:PQrb1}
\end{figure}

Inserting $M_b$ from (\ref{eq:Mb}) into the temperature yields the constraint  \begin{equation}
Q^2 + P^2 = \frac{\left(r_b^2-\chi ^2\right) \left(6 \chi  \left(r_b^2-\chi ^2\right)-l^2 \left(r_b-2 \kappa  \chi \right)\right)}{2 l^2 \chi }. \label{eq:rbk1}
\end{equation}
Replacing $P = - 2 i \chi g$, we can interpret this as a constraint on $r_b$. Then \eqref{eq:rbk1} has 4 solutions denoted  $r_{b,1\mp}$ and $r_{b,2\mp}$, each of which are  functions of $Q$, $g$ and $\chi$. We write these in the  appendix \ref{app:rbfunc}, and plot each as a function of $\chi$ in Fig.\ref{fig:PQrb1} for fixed values of $Q$ and $g$.

The quantities $r_{b,1+}$ and $r_{b,2+}$ are always positive, whereas  the positivity of $r_{b,1-}$ and $r_{b,2-}$  depends on the  parameter values; all
 positive values of  $r_{b,1-}$ yield negative entropy, but some positive values of $r_{b,2-}$ can yield positive entropy.  Hence we shall only consider $r_{b,1+}$, $r_{b,2+}$, and $r_{b,2-}$ in this section. 
 
 In the uncharged case as shown in Fig.~\ref{fig:EPQ0}, $r_b$ in the small TB-AdS started at $\chi=0$; upon turning on the electric charge $Q$ with the magnetic parameter $g=0$, the starting point is at a positive value of $\chi$, illustrated in the upper left plot in Fig.~\ref{fig:PQrb1}.
  As $Q$ increases the curve approaches an inflection point and retains the small/large branch shape for a while. Further increasing $Q$, the range of $r_b$ of the small TB-AdS gets shorter and disappears at around  $Q=0.13$.   
 However another solution of $r_{b,1+}$ at large values of $\chi$ emerges, extending the end point of $r_b$ of the large TB-AdS; this solution is denoted with the orange dashed line in the upper right plot in Fig.\ref{fig:PQrb1}.  All dashed lines in Fig.~\ref{fig:PQrb1} correspond to negative entropy, and so are not considered to be physical solutions. Thus as the electric charge $Q$ becomes larger than $\sim 0.13$ (with $g=0$) only the large (dyonic) TB-AdS solution exists.  
   
Once the magnetic parameter $g$ becomes nonzero, the bolt radius $r_b$ of the small TB-AdS  appears and again forms  the small/large branches, shown in
the lower  left plot in Fig.~\ref{fig:PQrb1}.   Setting $Q=0$, we recover the behaviour of the uncharged case; note that $P = - 2 i g \chi$ and so the pure magnetic case does not  exhibit behaviour similar to the pure electric case. If we increase $Q$ the small branch disappears (from the upper left to the upper right
plot in Fig.\ref{fig:PQrb1}, but if we increase $g$, then for $g > Q$ the small branch emerges again (the lower left plot becomes  the lower right one).

In general we find that   a maximum value of $\chi$ always exists. The dyonic TB-AdS solution has two branches (large and small) for two cases : $Q < 0.13$ with $g=0$, and $g > Q$.    Recall that a small (dyonic) TB-AdS is one whose radius increases as the NUT charge increases, whereas  the radius of a large (dyonic) TB-AdS decreases as the NUT charge increases.

\begin{figure}[b!]
\center{\includegraphics[scale=0.45]{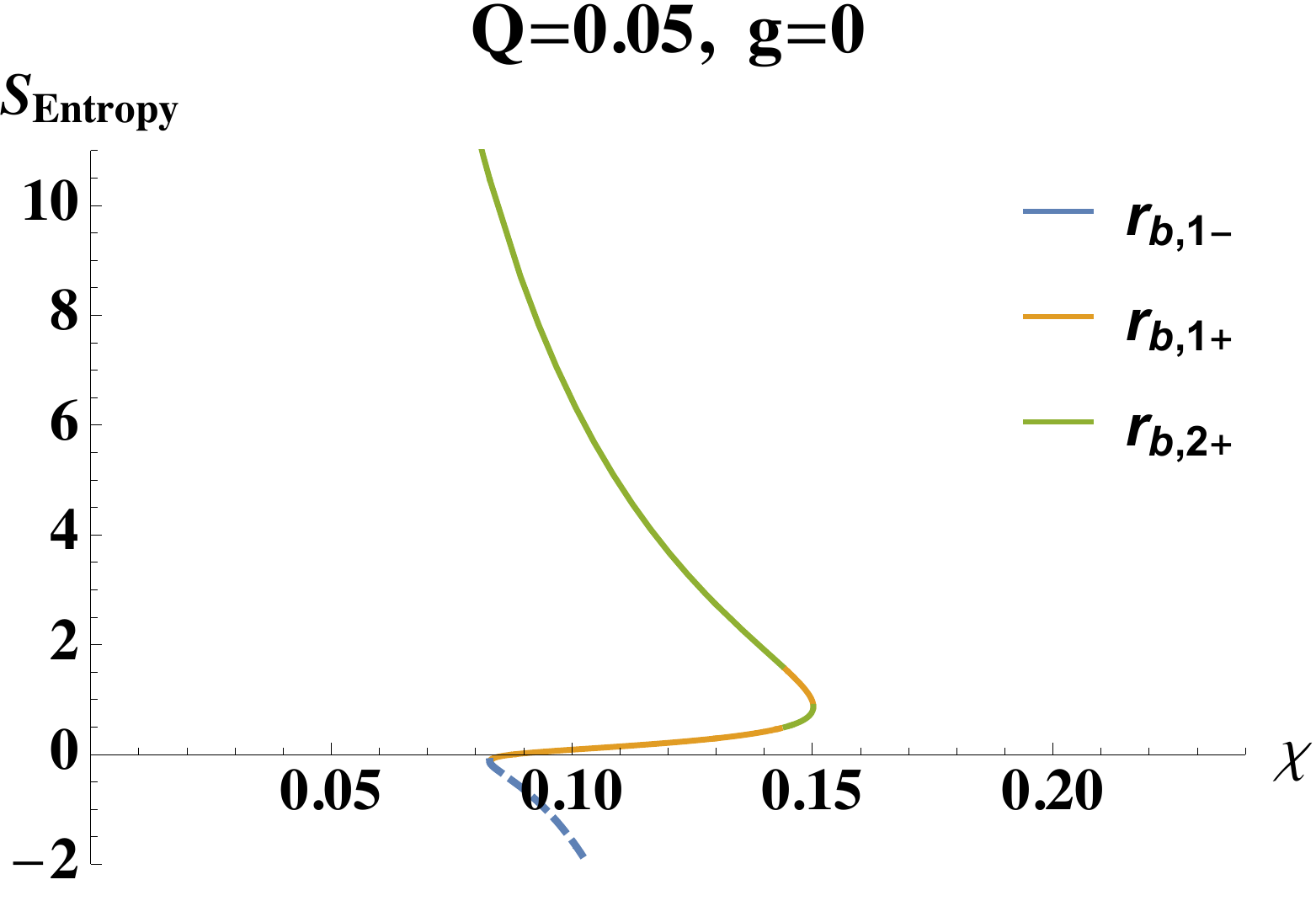} \; \; \includegraphics[scale=0.45]{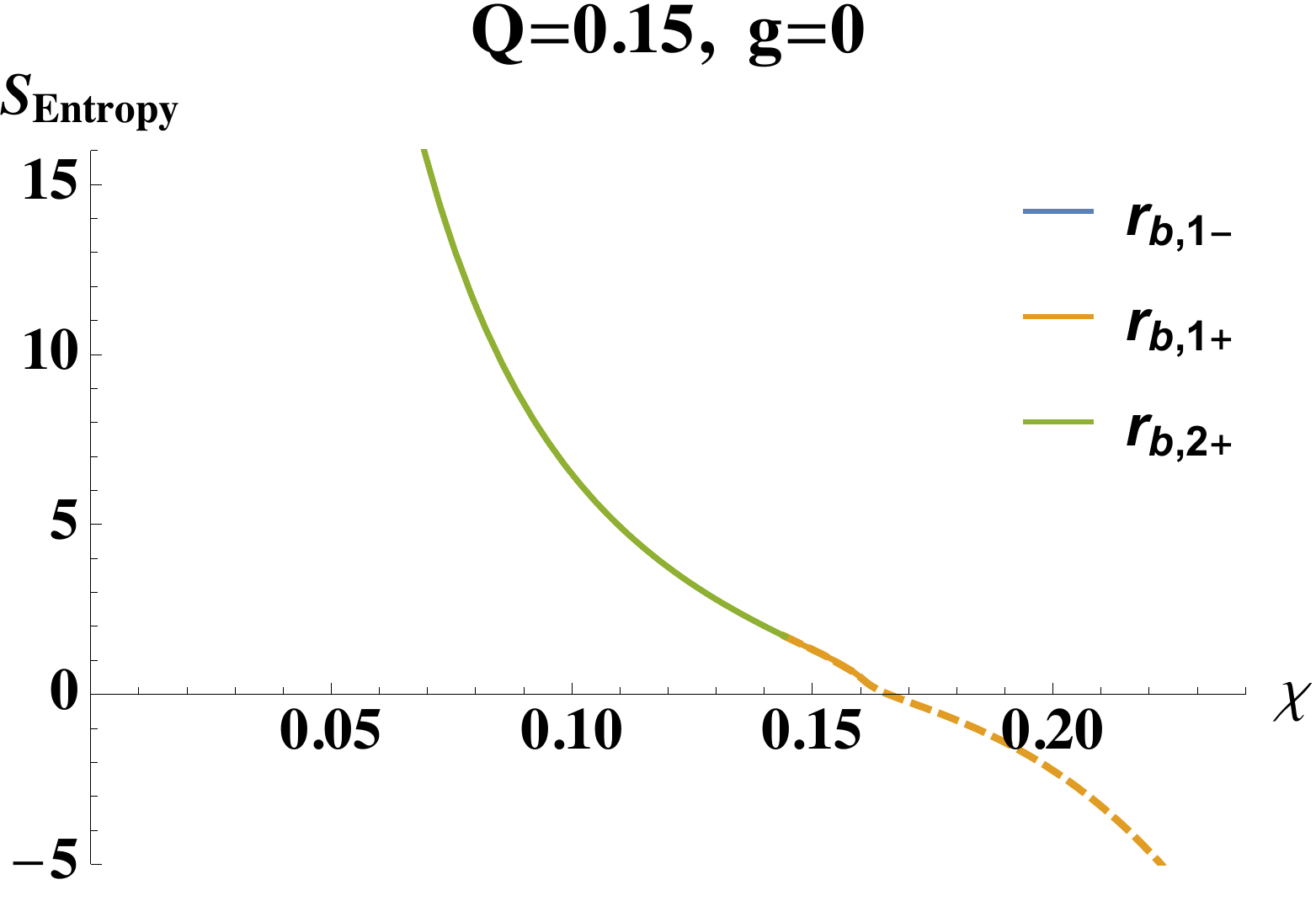} \\
\vspace{0.5cm}
\includegraphics[scale=0.45]{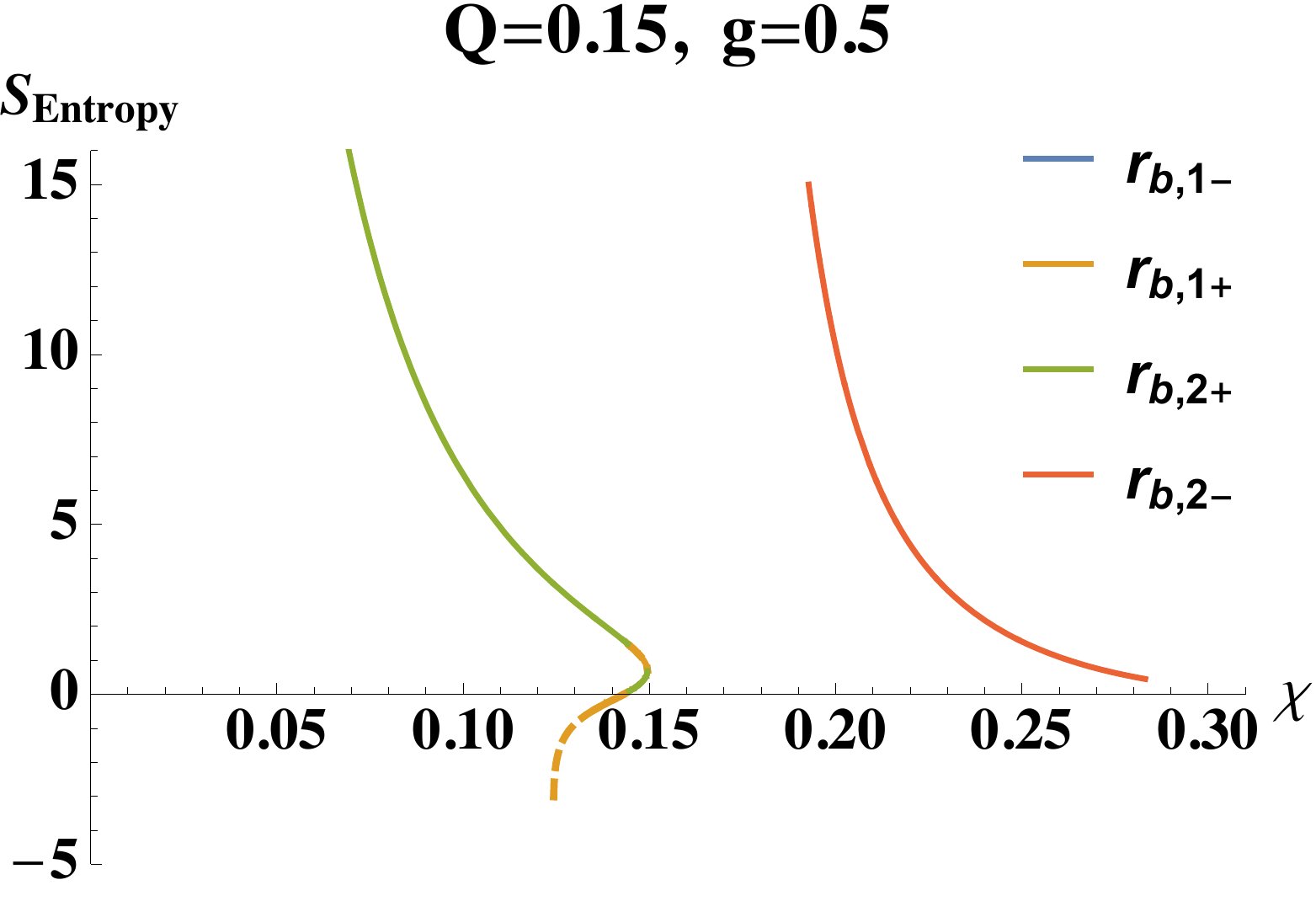} \; \; \includegraphics[scale=0.45]{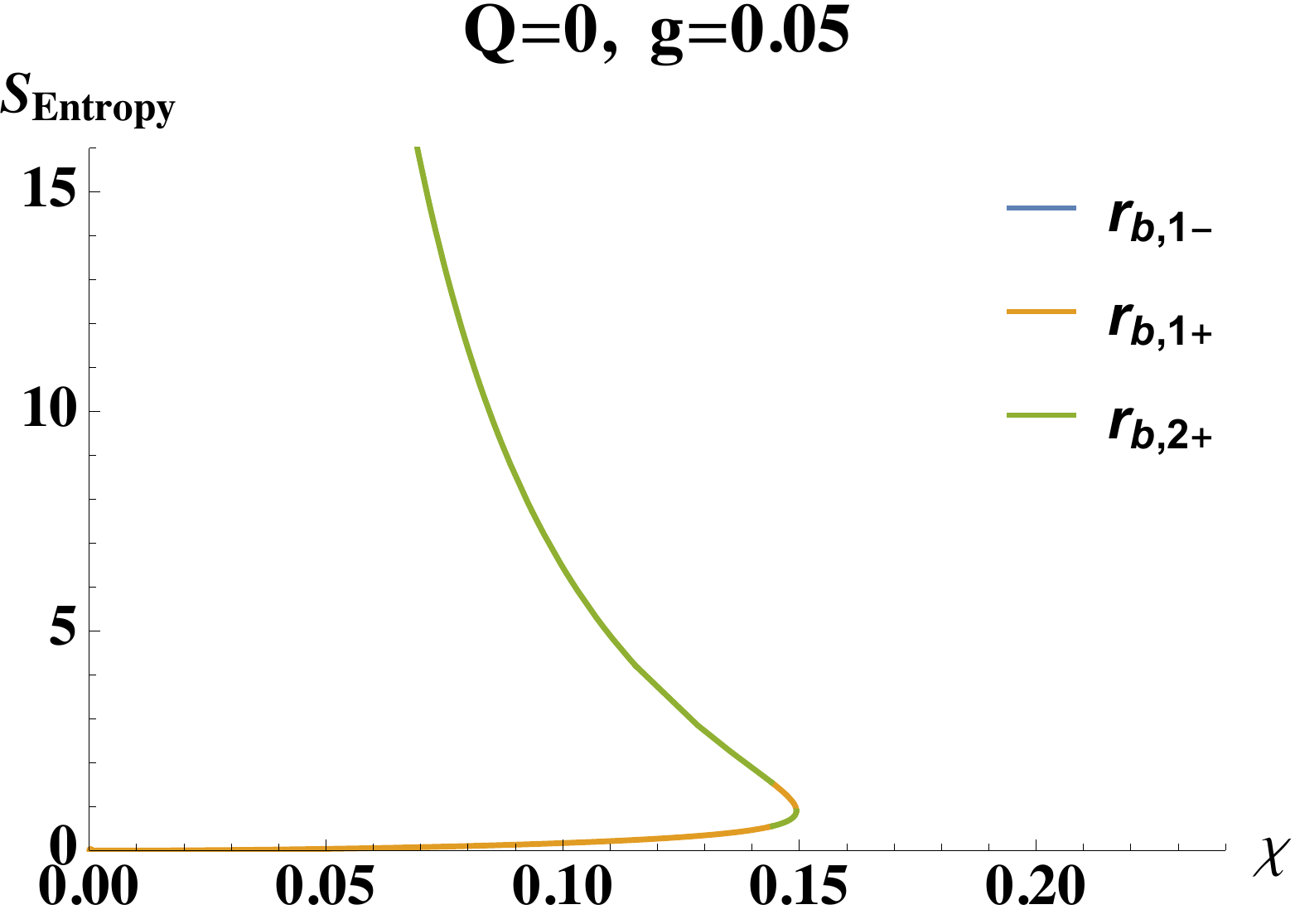}}
\caption{Entropy (\ref{eq:etnpTBk1}) of TB-AdS for $\kappa=1$ vs $\chi$ for fixed values of $Q$ and $g$ with $G=l=1$}
\label{fig:PQS}
\end{figure}

Inserting these values of $r_b$ into the entropy (\ref{eq:etnpTBk1}), we can express $S$ in terms of $Q$, $g$ and $\chi$. For  fixed values of $Q$ and $g$, the entropy as a function of $\chi$ is plotted in Fig.~{\ref{fig:PQS}}  where the dashed lines are ones calculated with the dashed lines of $r_{b,1+}$ in Fig.~\ref{fig:PQrb1}.  
 We see   that even for the $r_{b,+}$ solutions, the parameter range must be restricted to ensure positive entropy.  We also observe that if $Q$ is smaller than $0.13$ with $g=0$ or $g$ is bigger than $Q$, the behaviour of the entropy becomes similar to the uncharged TB-AdS case, shown in Fig.\ref{fig:SPQ0}, but as $Q$ increases keeping $g=0$ the  only allowed solution is that of the large (dyonic) TB-AdS.

\begin{figure}[b!]
\center{\includegraphics[scale=0.45]{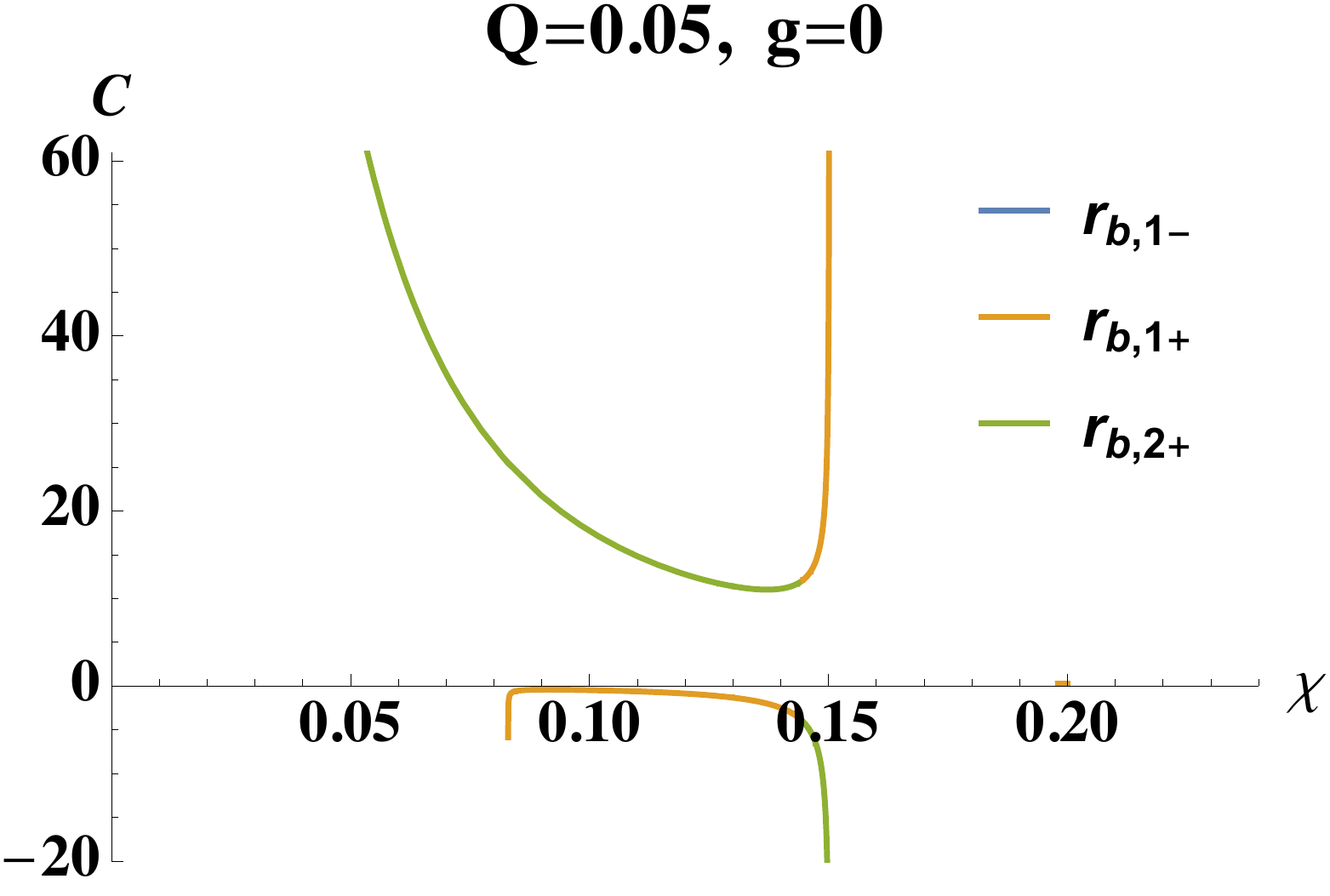} \; \; \includegraphics[scale=0.45]{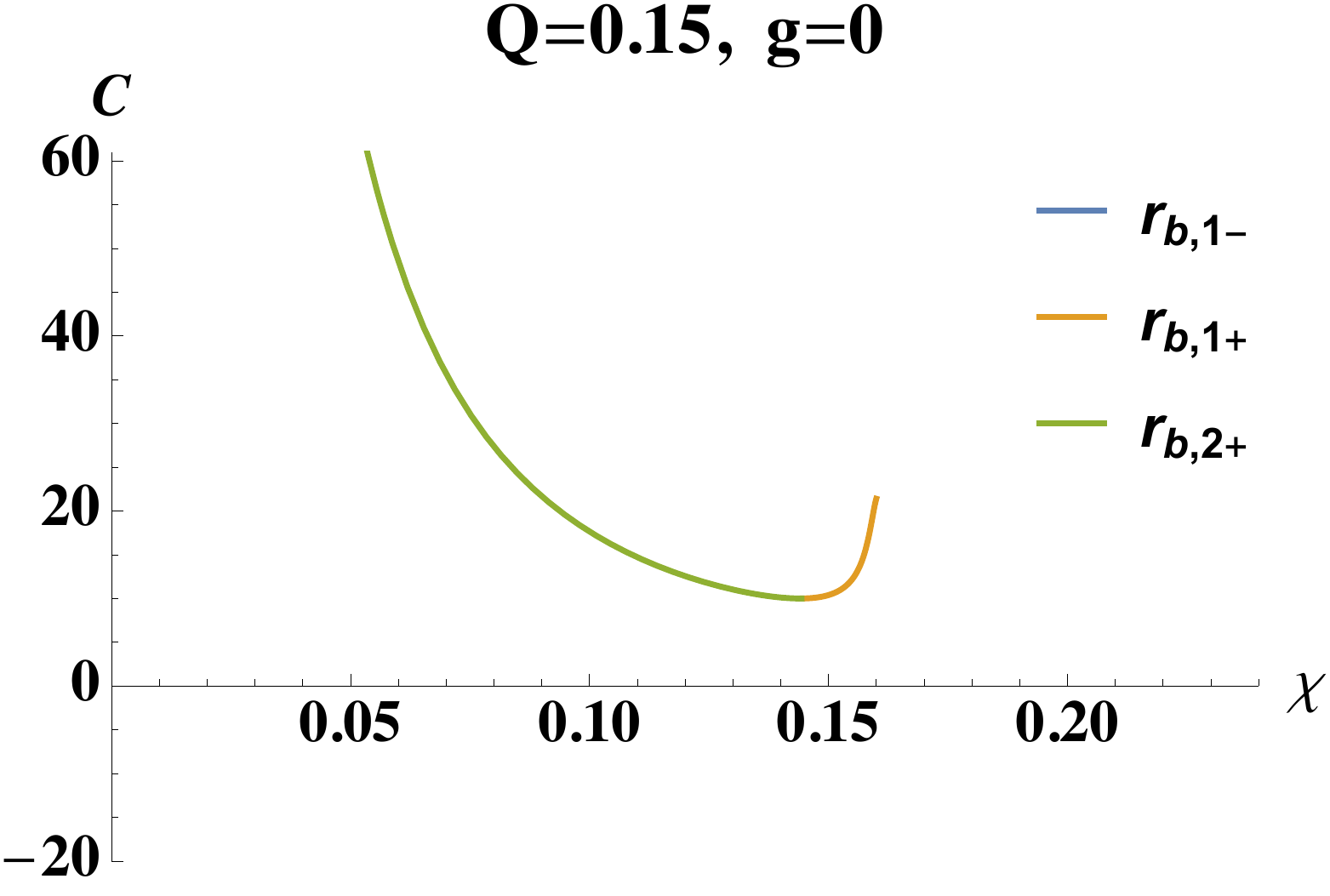} \\
\vspace{0.5cm}
\includegraphics[scale=0.45]{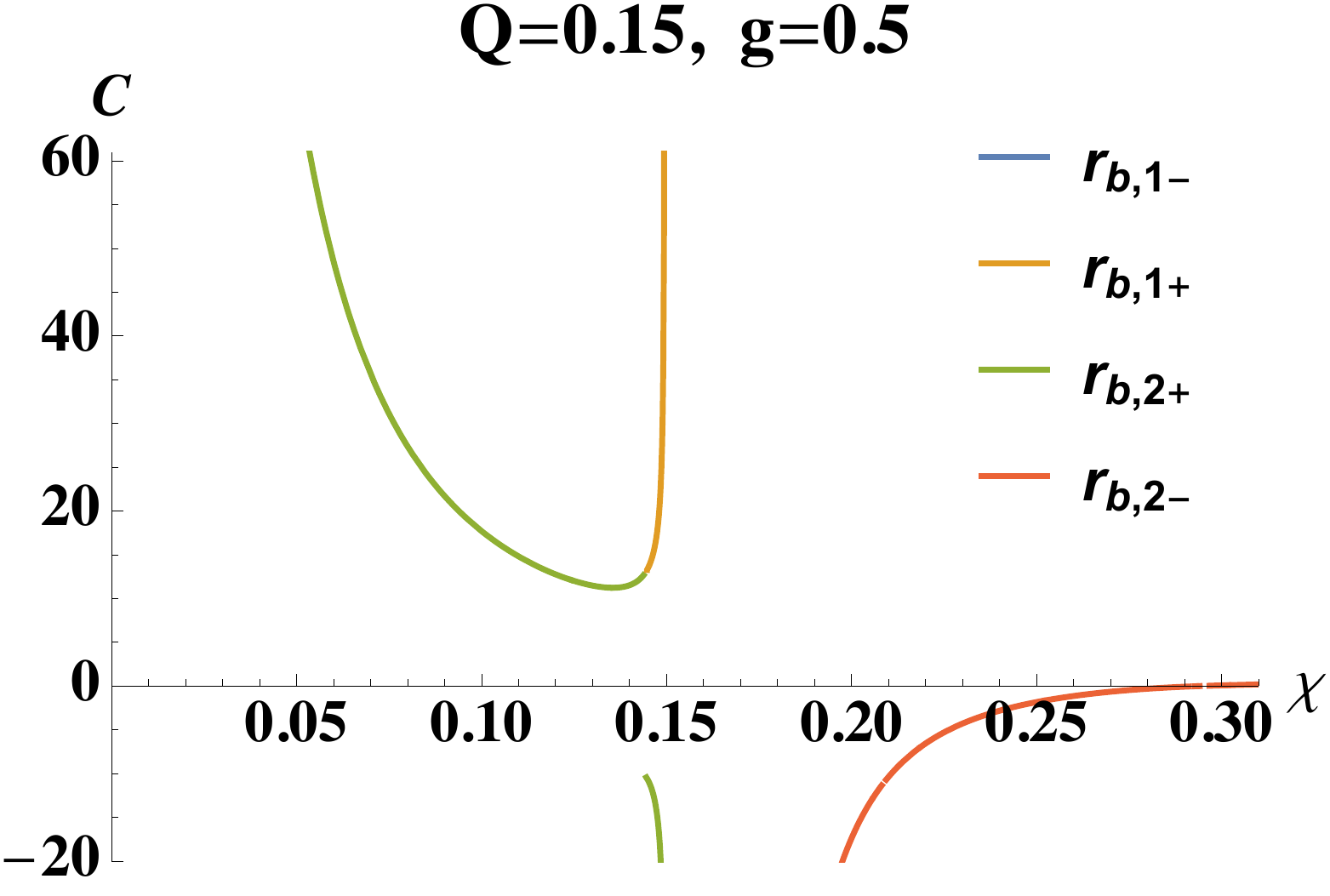} \; \; \includegraphics[scale=0.45]{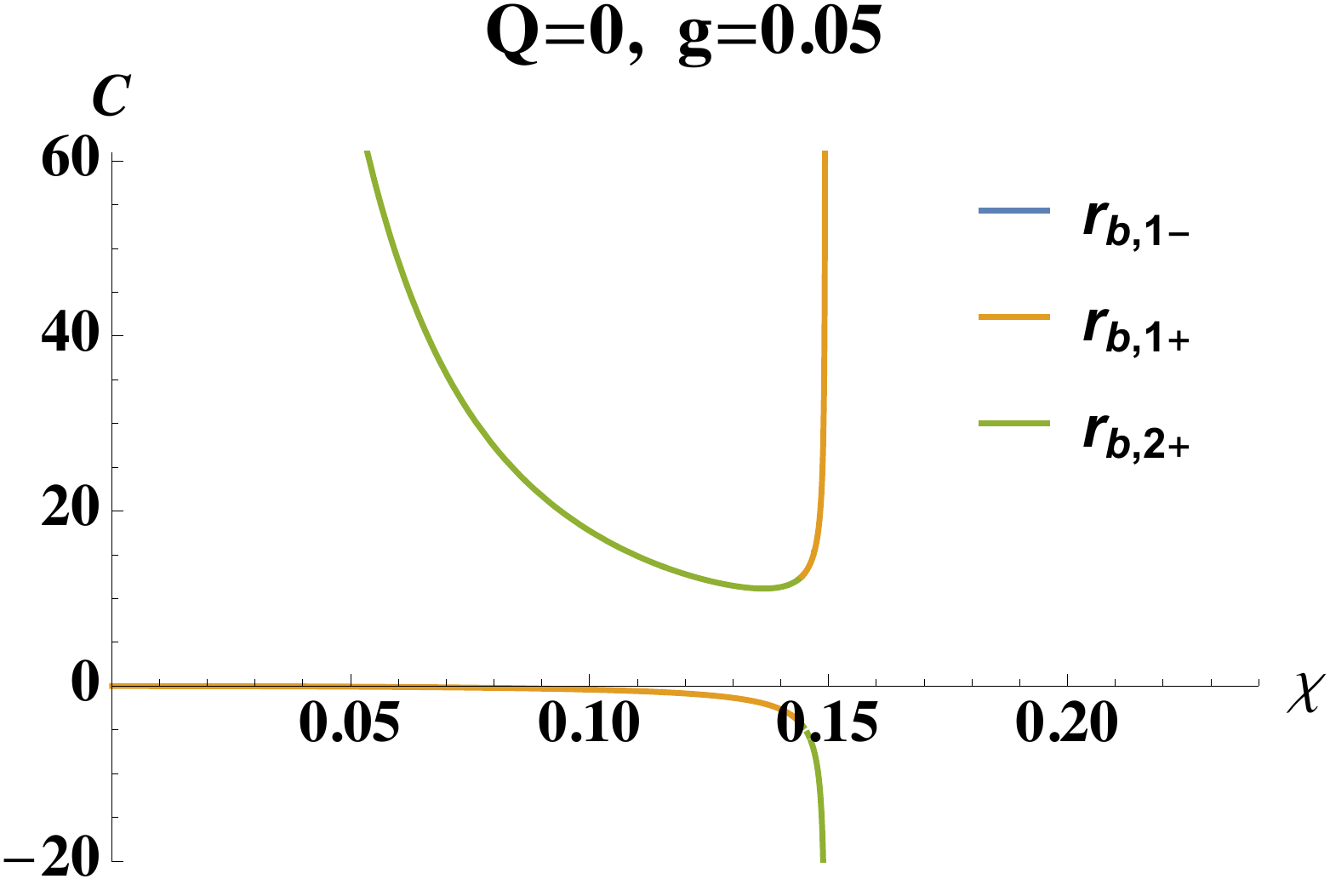}}
\caption{Heat capacity (\ref{eq:Ck1}) for TB-AdS for $\kappa=1$ vs $\chi$ for fixed $Q$ and $g$ with  $G=l=1$}
\label{fig:PQC}
\end{figure}

From the definition (\ref{eq:C}), we also calculate the heat capacity 
\begin{align}
C_{P, Q} =& \frac{\omega l^2 r_b \chi^2}{2 G a_1}  \bigg(\frac{\left(r_b^2-\chi^2\right)^2 \left(l^2 r_b-12 \chi^3\right)}{l^2 \chi^2} \nonumber\\
&+\frac{4 \chi \left(Q^2-P^2\right) \left(9 r_b^4+14 r_b^2 \chi^2+\chi^4\right)+16 i P Q r_b \left(r_b^4+8 r_b^2 \chi^2+3 \chi^4\right)}{\left(r_b^2-\chi^2\right)^2}\bigg) \label{eq:Ck1}
\end{align}
for  fixed $P$ and $Q$, where 
\begin{equation}
a_1 = 12 \chi  r_b \left(r_b^2-\chi ^2\right)^2 + l^2 \left(2 \chi ^2 r_b^2 - r_b^4 - \chi ^4  + 4 \chi  r_b \left(P^2+Q^2\right) \right).
\end{equation}
Substituting $r_{b,1+}$ and $r_{b,2+}$, we plot the heat capacity (\ref{eq:Ck1}) as a function of $\chi$ in Fig.~{\ref{fig:PQC}}.  If $Q < 0.13$ with $g=0$ or as $g$ becomes  larger than $Q$, 
the heat capacity exhibits behaviour similar  to the uncharged TB-AdS case, shown  in Fig.\ref{fig:CPQ0}; the small (dyonic) TB-AdS has negative heat capacity and the large (dyonic) TB-AdS has positive heat capacity. Other cases, for which $r_{b,1+}$ and $r_{b,2+}$ are restricted to yield positive entropy,  have only positive heat capacity, shown in the upper right graph of
Fig.~\ref{fig:PQC}. 

From the Euclidean action, the energy, electric, and magnetic potentials are obtained from (\ref{eq:emp1}) and (\ref{eq:defE})
\begin{align}
&\Phi_{\textrm{E}}^{(1)} = \frac{\omega}{4 \pi G} \frac{r_b \left(2 i P \chi  r_b+Q (r_b^2+ \chi ^2) \right)}{\left(\chi ^2-r_b^2\right)^2}, \qquad \Phi_{\textrm{M}}^{(1)}= -\frac{\omega}{4 \pi G}  \frac{r_b \left(P (r_b^2+ \chi ^2) - 2 i Q \chi  r_b\right)}{ \left(\chi ^2-r_b^2\right)^2}. \label{eq:empk1L} \\
&E^{(1)} =\frac{\omega}{8 \pi G}  \left[M_b +\frac{r_b^2+\chi ^2}{4 \chi }-\frac{3 \chi ^2 r_b+r_b^3}{l^2}+\frac{\left(Q^2-P^2\right) \left(-3 \chi ^4 r_b-6 \chi ^2 r_b^3+r_b^5\right)}{\left(r_b^2-\chi ^2\right)^3}+\frac{16 i P Q \chi ^3 r_b^2}{\left(\chi ^2-r_b^2\right)^3}\right] , \label{eq:ETBk1em1} 
\end{align}
and by  (\ref{eq:emp2}) and (\ref{eq:defE})
\begin{align}
\Phi_{\textrm{E}}^{(2)}= &\frac{\omega}{4 \pi G} \frac{(Q r_b +i P \chi)}{\left(r_b^2-\chi^2\right)}, \qquad \Phi_{\textrm{M}}^{(2)}= \frac{\omega}{4 \pi G }\frac{(-P r_b+i Q \chi)}{\left(r_b^2-\chi^2\right)}.\label{eq:empk1Ls} \\
E^{(2)} =& \frac{\omega}{8 \pi G}  \bigg[ M_b +\frac{r_b^2+\chi^2}{4 \chi} -\frac{r_b \left(r_b^2+3 \chi^2\right)}{l^2}+\frac{1}{(r_b^2-\chi^2)^3} \bigg( P^2 r_b (r_b^4+6 r_b^2 \chi^2+\chi^4 ) \nonumber\\
&+2 i P Q \chi (-3 r_b^4-6 r_b^2 \chi^2+\chi^4 )+Q^2 r_b (r_b^4-10 r_b^2 \chi^2+\chi^4) \bigg) \bigg]. \label{eq:ETBk1ems} 
\end{align}

As stated above, we find that the first law cannot be satisfied unless the regularity condition (\ref{eq:regcd1}) is imposed. Upon doing this, we find 
that  (\ref{eq:ETBk1em1}) - (\ref{eq:empk1Ls}) and (\ref{eq:etnpTBk1}) become
\begin{align}
&E = \frac{\omega}{8 \pi G}  \left[M_b + \frac{r_b^2+\chi ^2}{4 \chi } -\frac{r_b \left(r_b^2+3 \chi ^2\right)}{l^2}+\frac{Q^2 \left(3 \chi ^2 r_b+r_b^3\right)}{\left(r_b^2+\chi ^2\right)^2}\right],  \label{eq:Ek1RC}\\
&S =\frac{\omega}{4 G}  \left[r_b^2+\chi ^2 -\frac{24 \chi ^3 r_b}{l^2} + \frac{8 Q^2 \chi ^3 r_b}{\left(r_b^2+\chi ^2\right)^2}\right], \label{eq:Sk1RC}\\
&\Phi_{\textrm{E}} = \frac{\omega}{4 \pi G}\frac{Q r_b}{\left(r_b^2+\chi ^2\right)}, \qquad  \Phi_{\textrm{M}} = 0 
\label{eq:EPk1RC}
\end{align} 
and that these quantities satisfy the first law
\begin{align}
&dE = T d S + \Phi_{\textrm{E}} dQ\\
&F = E - T S - \Phi_{\textrm{E}} Q 
\label{freeen}
\end{align}
with the  free energy written as in \eqref{freeen}.  Note that $g$ and $Q$ are no longer independent.

\section{Zero temperature limit of dyonic TB-AdS}

For vanishing Hawking temperature  the TB-AdS metric (\ref{eq:TNgE})  factorizes:
\begin{align}\label{fEext}
f_{\textrm{E}} =  \frac{1}{l^2(r^2-\chi^2)} (r+r_0 + \alpha)(r+r_0 - \alpha)(r - r_0)^2
\end{align}
where
\begin{align}
&r_0 = \sqrt{\chi ^2 + \frac{1}{6} l \left( \sqrt{\kappa ^2 l^2+12 \left(Q^2+P^2\right)}-\kappa  l\right)}, \\
&\alpha = \sqrt{4 \chi ^2-\frac{1}{3} l \left(\sqrt{\kappa ^2 l^2+12 \left(Q^2+P^2\right)}+2 \kappa  l\right)}.
\end{align}
By writing 
\begin{align}
\tau \rightarrow \frac{\tilde{\tau}}{\epsilon}, \qquad r \rightarrow r_0 + \epsilon \tilde{r}
\end{align}
the metric   (\ref{eq:elds}) with $f_E$ given by \eqref{fEext} becomes
\begin{equation}\label{AdS2Mk-eps}
ds_{\textrm{ext}}^2 \sim \frac{\tilde{r}^2 f_{\epsilon,-}}{l^2 (r_0^2-x^2)^2} (d \tilde{\tau} + 2 \chi \lambda(\theta) \epsilon d\phi)^2 + \frac{l^2 r_0 f_{\epsilon,+}}{\tilde{r}^2 \left(4 r_0^2- \alpha ^2\right)^2} d\tilde{r}^2 + (r_0^2- \chi^2 +2 \tilde{r} r_0 \epsilon) d \Omega_{\kappa}, 
\end{equation}
to leading order in $\epsilon$.  In the limit $\epsilon \rightarrow 0$ we obtain the near-horizon metric
\begin{align}\label{AdS2Mk}
ds_{\textrm{ext}}^2 \to \frac{\tilde{r}^2}{l^2}\frac{(4 r_0^2 - \alpha^2)}{(r_0^2 - \chi^2)} d\tilde{\tau}^2 + \frac{l^2}{\tilde{r}^2} \frac{(r_0^2 - \chi^2)}{(4 r_0^2 - \alpha^2)} d \tilde{r}^2 + (r_0^2 - \chi^2) (d \theta^2 + Y(\theta)^2 d \phi^2) 
\end{align}
which is  $AdS_2 \times \mathcal{M}_{\kappa}$, where $\mathcal{M}_{\kappa}$ is a two dimensional manifold depending on $\kappa$ as given by \eqref{mankappa}, and   where
\begin{align}
&4 r_0^2 - \alpha^2 = l \sqrt{\kappa^2 l^2 + 12(Q^2 + P^2)} \nonumber \\
&r_0^2 - \chi^2 = \frac{l}{6} \sqrt{\kappa^2 l^2 + 12(Q^2 + P^2)} - \frac{\kappa}{6} l^2
\label{AdS2-relns}
\end{align}
It is interesting to note that no NUT charge $\chi$ appears in the $AdS_2$ section due to the relations \eqref{AdS2-relns}, and the metric \eqref{AdS2Mk} has no singularities at the horizon.  

 Thus for $\kappa \neq 1$ the extremal limit is given by taking the metric function to be (\ref{fEext}) and we can check if the thermodynamic relations hold  in this limit. The Hawking temperature  \eqref{eq:etnpTBkn2} is
\begin{align}
T_{\textrm{H}} &= \frac{3(r_b^2 - \chi^2)^2+ l^2 (-Q^2 - P^2+ (r_b^2- \chi^2) \kappa)}{4 \pi r_b l^2(r_b^2 - \chi^2)} 
\label{eq:HkT}
\end{align}
upon substituting for the mass parameter. At  zero temperature the following condition 
\begin{equation}
Q_{ext}^2 + P_{ext}^2 = \frac{\left(r_b^2-\chi ^2\right) \left(3 r_b^2-3 \chi ^2+\kappa  l^2\right)}{l^2} \label{eq:PQ2ext}
\end{equation}
must hold, implying that  
\begin{align}
M &=M_{ext}  = \frac{r_b \left(2 r_b^2-6 \chi ^2+\kappa  l^2\right)}{l^2} \nonumber \\
&= \frac{\left(2 \kappa  l^2-12 \chi ^2+l \sqrt{\kappa ^2 l^2+12 (Q_{ext}^2+P_{ext}^2)}\right) \sqrt{6 \chi ^2-l^2 \kappa  + l \sqrt{\kappa ^2 l^2+ 12(Q_{ext}^2+P_{ext}^2)}}}{3 \sqrt{6} l^2}.
\label{eq:exM}
\end{align}
as well as
\begin{equation}
Q_{ext}^2 = \left(r_b^2+\chi ^2\right){}^2 \left(\frac{\kappa }{r_b^2-\chi ^2}+\frac{3}{l^2}\right) 
\label{eq:Qext}
\end{equation}
upon imposing the regularity condition. 

The free energy and the energy (\ref{eq:Ekn1}) become
\begin{equation}
F_{ext} = -\frac{\omega r_b}{4 \pi G} \left(\frac{\kappa  \chi ^2}{r_b^2-\chi ^2} + \frac{r_b^2+3 \chi ^2}{l^2}\right), \qquad E_{ext} =\frac{\omega r_b^3}{4 \pi G} \left(\frac{\kappa }{r_b^2-\chi ^2}+\frac{2}{l^2}\right). \label{eq:FEext}
\end{equation}
This satisfies the following relation
\begin{align}
&F_{ext} = E_{ext} - \Phi_{\textrm{E}} Q |_{ext},  \label{eq:Fext} \\
&dE_{ext} =  \Phi_{\textrm{E}} d Q |_{ext} \label{eq:FLext}
\end{align}
where we used $\Phi_{\textrm{E}}$ in (\ref{eq:EPk1RC}) with $Q=Q_{ext}$
\begin{equation}
\Phi_{\textrm{E}} = \frac{\omega}{4 \pi G}\frac{Q_{ext} r_b}{\left(r_b^2+\chi ^2\right)}. \label{eq:Phiext}
\end{equation}
As there is no Misner string, we expect that the entropy is still given by  (\ref{eq:etnpTBkn1}) 
\begin{equation}
S_{ext} =    \frac{\omega l \left( \sqrt{\kappa^2 l^2 + 12(Q^2 + P^2)} -  \kappa  l\right)}{24 G}
\end{equation}
which is the Bekenstein-Hawking entropy.

For $\kappa=1$ the situation is more complicated. If  the Hawking temperature (\ref{eq:TTB}) is identified with the time periodicity $8 \pi \chi$ needed to eliminate the Misner string, then the  zero temperature limit is approached by taking $\chi \rightarrow \infty$, which  would lead us to conclude there is no zero temperature limit for the TB-AdS solution. The alternative is to abandon this identification in the extremal case since the Misner string does not appear in the near-horizon limit and imposing   time periodicity cannot be done due to the long throat of  $AdS_2$ ;  the Bekenstein-Hawking entropy formula is then restored even for $\kappa=1$ and we find the above thermodynamic relations hold for $\kappa=1$ as well.

\begin{figure}[t!]
\center{\includegraphics[scale=0.45]{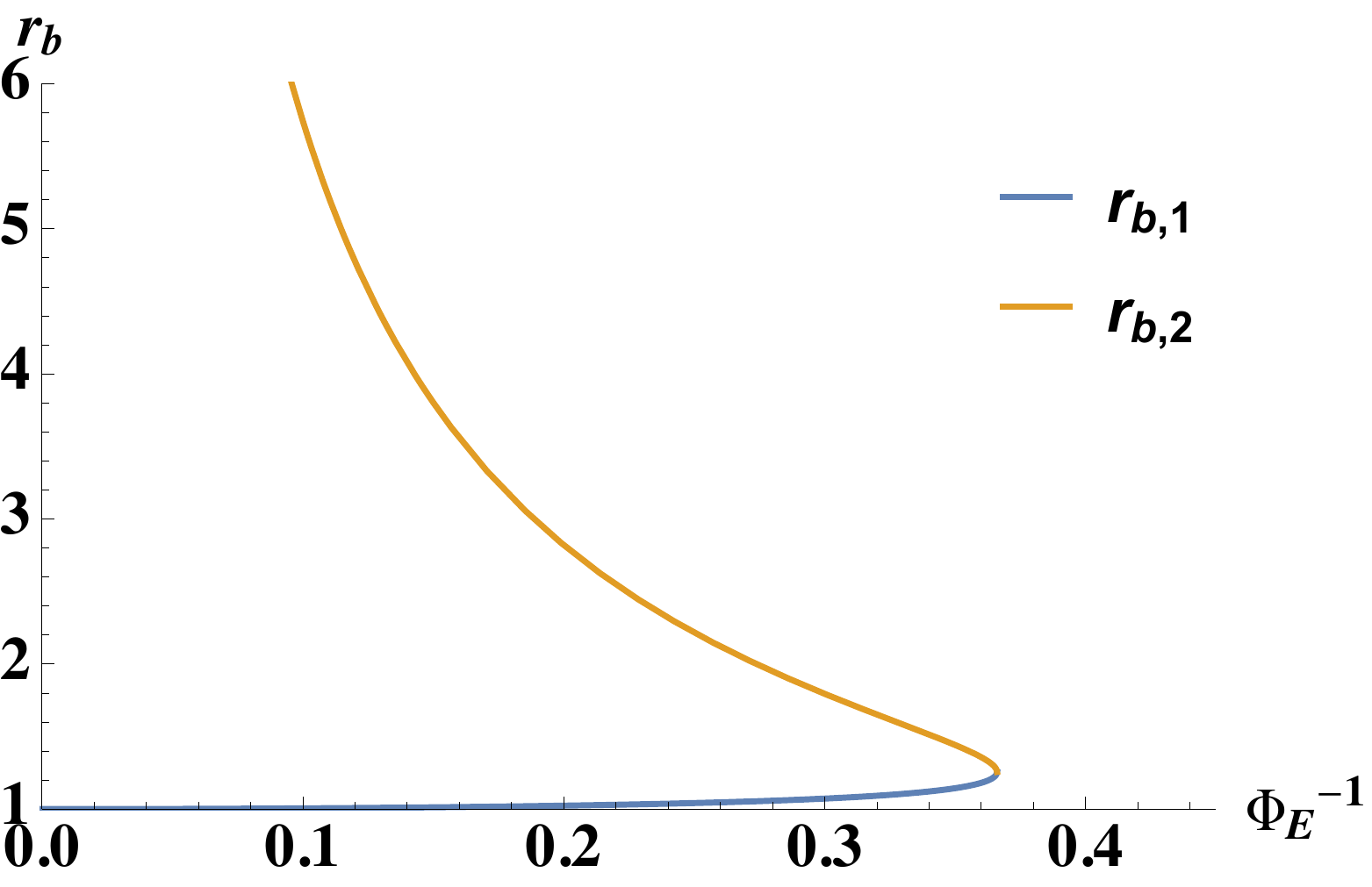} $\qquad$ \includegraphics[scale=0.45]{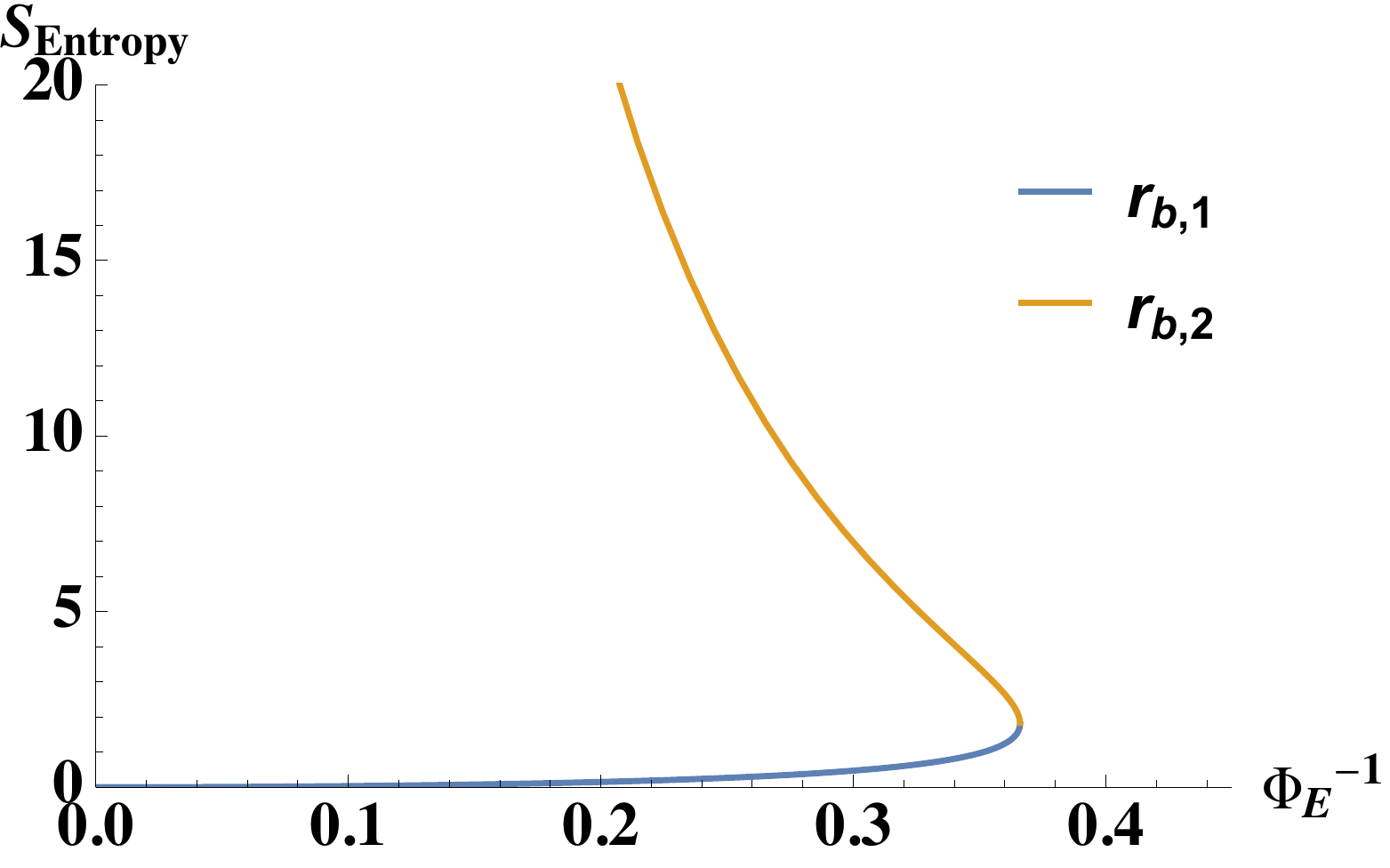}}
\caption{Left: TB-AdS radius vs ${\Phi_E}^{-1}$. Right: entropy vs  ${\Phi_E}^{-1}$  for the extremal TB-AdS for a fixed value of $\chi$ and $\kappa=1$ with $G=l=1$}
\label{fig:ExtrbnS}
\end{figure}

Continuing with this latter assumption, we can observe new ``finite temperature-like" behaviour as follows.
Inspired by the idea in \cite{Choi:2018vbz}, let us identify the electric potential as a ``temperature-like'' quantity in this extremal limit $T_{\textrm{H}} \rightarrow 0$. 
 Solving (\ref{eq:Phiext}) for $Q_{ext}$ and inserting this into \eqref{eq:Qext} yields
\begin{equation}
r_{b\pm} = \sqrt{\frac{\chi ^2}{2}-\frac{\kappa  l^2}{6} +\frac{8 \pi^2 G^2 l^2}{3 \Phi_E^2 \omega ^2} \mp  \frac{\pi}{3 \omega} \sqrt{\left( \frac{\omega ^2}{4 \pi^2} \left(\kappa  l^2-3 \chi ^2\right)-\frac{4 G^2 l^2}{\Phi_E^2}\right)^2-\frac{12 G^2 l^2 \chi ^2 \omega ^2}{\pi^2 \Phi_E^2}}  }
\end{equation}
as the physically acceptable values of  $r_b$ in terms of  ${\Phi_E}^{-1}$ and the other parameters. We plot $r_{b\pm} $ as a function of 
${\Phi_E}^{-1}$  in  Fig.\ref{fig:ExtrbnS} where $r_{b,1}=r_{b-}$  and $r_{b,2} = r_{b+}$.  The resemblance to the uncharged TB-AdS behaviour displayed in Fig.\ref{fig:EPQ0} and Fig.\ref{fig:SPQ0} is clear, with the notable exception that the $x$-axis is the inverse of an electric potential and not of temperature. 
\begin{figure}[t!]
\center{ \includegraphics[scale=0.5]{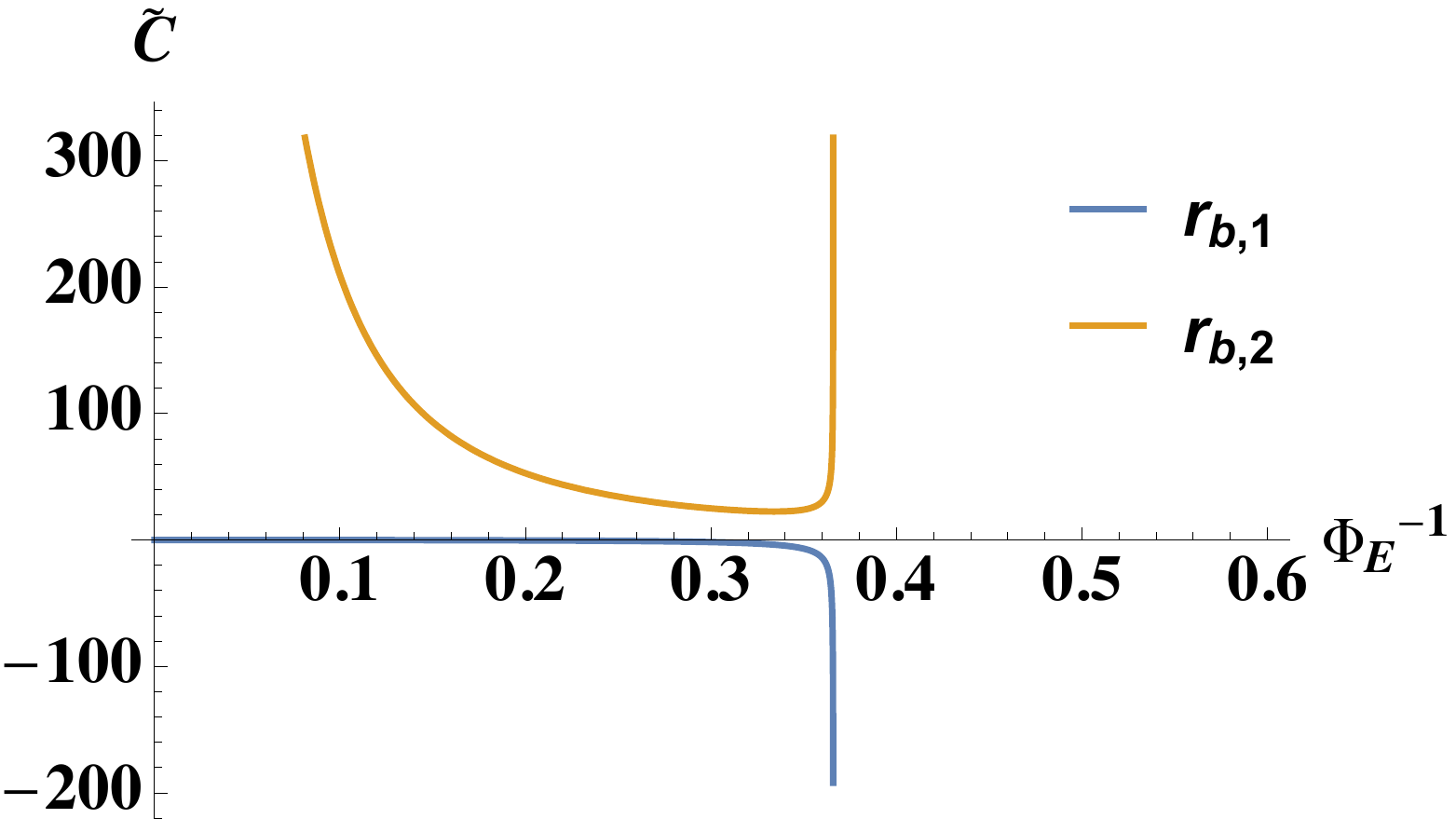} }
\caption{``Heat capacity-like" quantity vs ${\Phi_E}^{-1}$ for the extremal TB-AdS with a fixed $\chi$ and $\kappa=1$}
\label{fig:ExtC}
\end{figure}

We can  calculate a ``heat capacity-like" quantity  
\begin{equation}
C = T \frac{\partial S}{\partial T}\bigg|_{Q,P} \qquad \rightarrow \qquad \tilde{C} = \Phi_{E} \frac{\partial S}{\partial \Phi_{E}}
\end{equation}
replacing temperature with  the inverse of  the electric potential.  We obtain
\begin{equation}
\tilde{C} = \frac{\omega  r_b^2 \left(r_b^2-\chi ^2\right) \left(3 r_b^2+\kappa  l^2-3 \chi ^2\right)}{2 G \left(-6 \chi ^2 r_b^2+3 r_b^4-\kappa  l^2 \chi ^2+3 \chi ^4\right)}. 
\end{equation}
and display in  Fig.~{\ref{fig:ExtC}}  the behaviour of $\tilde{C}$ as a function of ${\Phi_E}^{-1}$ .  We observe  that the behaviour is  similar to that of  the uncharged TB-AdS case.

\section{Conclusion}

We have investigated thermodynamics for  dyonic Euclidean Taub-NUT-AdS spacetime, extending the previous studies \cite{Chamblin:1998pz,Johnson:2014xza,Johnson:2014pwa}.  

After reviewing the  thermodynamics of the uncharged TN/TB-AdS black holes, we investigated their dyonic charged counterparts.  
For the charged TN-AdS, the condition $P=iQ$   is always necessary and so there is no distinction between electric and magnetic charges. For the dyonic TB-AdS, we  found that the regularity condition  (\ref{eq:regcd1}) again made the magnetic charge dependent on the electric charge.  This condition was necessary in order that the first law of thermodynamics hold.  In this case, both the charged TN/TB-AdS satisfy
\begin{equation*}
dE = T dS + \Phi_E dQ, \qquad F = E - T S - \Phi_E Q
\end{equation*}
and where  the entropy $S$ is the Bekenstein-Hawking entropy for TB-AdS provided $\kappa \neq 1$; in this case both $r_b$ and $Q$ are independent parameters. For $\kappa=1$  the entropy $S$ does not follow the Bekenstein-Hawking formula due to the additional   contribution of the Misner string, and because of requiring the time periodicity the phase space is reduced again. Thus the variation in the first law is made by only one variable such as $Q$. 
Furthermore, we explored the thermodynamic behaviours of entropy and heat capacity before imposing the regularity condition for $\kappa=1$ case. There is a maximum NUT charge, and small/large TB-AdS phases, whose radii increase/decrease as the NUT charge increases, whose  heat capacity is negative/positive respectively. These characteristics appear  either when the magnetic parameter is turned off and the electric charge is smaller than $0.13$ ($l=1$ and $G=1$) or when the magnetic parameter becomes larger than twice the electric charge. For other parameter ranges, the behaviour is like that of the large TB-AdS.

Under this configuration we studied the zero temperature limit of TB-AdS. We showed that the near horizon geometry becomes $AdS_2 \times \mathcal{M}_{\kappa}$, where $\mathcal{M}_{\kappa}$ is a two dimensional manifold depending on $\kappa$. This indicates that the Misner string disappears at the horizon in this limit. 
Retaining the standard $1/8\pi\chi$ time periodicity employed at nonzero temperature implies that    $\chi \rightarrow \infty$ in the the zero temperature limit, 
in contradiction with \eqref{eq:HkT} which would force $T\to\infty$. This contradiction indicates that there might not exist zero temperature limit in this case.  However as the near horizon geometry of the extremal case is $AdS_2 \times \mathcal{M}_{\kappa}$ (with no Misner string at the horizon), we note that  time periodicity  cannot be imposed on $AdS_2$ in this limit, suggesting we can relax this condition.  So doing, we find that we can make sense of the thermodynamics in the
 zero temperature limit 
 and speculate the Bekenstin-Hawking entropy formula  holds even for $\kappa=1$ case. Furthermore, we showed that  inverse of the electric potential can be regarded as a ``temperature-like" quantity~\cite{Choi:2018vbz}, with the zero-temperature  TB-AdS  having  entropy  and  a ``heat capacity-like''  quantity exhibiting similar behaviour to the the heat capacity of the uncharged TB-AdS solution at finite temperature. The full implications of abandoning time periodicity in the extremal case for $\kappa=1$ remain to be understood.


Since Taub-NUT-AdS is asymptotically locally AdS spacetime, it has been drawn attention along with asymptotically AdS spacetime in AdS/CFT perspective \cite{Chamblin:1998pz,Martelli:2012sz,Toldo:2017qsh}. The dual field theory of Taub-NUT spacetime is known to be a conformal theory on a squashed sphere. It would be interesting if we find that entropy calculated at the boundary field theory recovers the gravity results under both the regularity conditions of a gauge field.

\appendix

\section{Thermodynamics of the Dyonic RN black hole \label{sec:appRNbh}}

Here we summarize thermodynamics of dyonic RN black hole \cite{Lee:2018hrd} and show that $\Phi_{\textrm{E}}^{(1)}$ and $\Phi_{\textrm{E}}^{(2)}$ yield  equivalent results.
 The dyonic RN black hole metric has the form
\begin{equation}
ds^2 = - f(r) dt^2 + \frac{1}{f(r)} dr^2 + dr^2 (d\theta^2 + \sin(\theta)^2 d \phi^2) ,
\end{equation}
with
\begin{equation}
f(r) = 1 - \frac{2 M}{r} + \frac{q^2 + p^2}{4 r^2}.
\end{equation}
The gauge field and its corresponding field strength become
\begin{equation}
A = \frac{1}{\kappa} \bigg(\frac{q}{r} -\frac{q}{r_h} \bigg) dt - \frac{p}{\kappa} \cos \theta d \phi, \qquad F = \frac{1}{\kappa} \bigg(\frac{q}{r^2} dt \wedge dr + p \Omega_2 \bigg) 
\end{equation}
where we fixed the gauge for $A$ to be regular at the horizon and then its norm is written as
\begin{equation}
A^2(r_h) = \frac{1}{\kappa^2} \frac{p^2}{r_h^2} \cot^2 (\theta).
\end{equation}
The divergence that occurs at $\theta=0$ and $\pi$ in $A^2$ can be removed by defining the gauge field differently in the coordinate patches covering
 $\theta=0$ and $\pi$ so that it is regular on both. The gauge fields can be matched up to a gauge transformation where the patches overlap. All thermodynamic quantities can be calculated in a quasilocal frame  \cite{Lee:2018hrd}, with the limit $R \rightarrow \infty$ taken at the end. The free energy from the semi classical approximation is then
\begin{align}
&I_{\textrm{ren}} = \beta F = -\frac{4 \pi ^2 r_h \left(4 M r_h+p^2-q^2\right)}{\kappa ^2 \left(M-r_h\right)} 
\end{align}
which is equivalent to
\begin{equation}
F = E - TS - \Phi_{\textrm{E}}Q_E \label{app:I_ren}
\end{equation}
and where the various thermodynamic quantities are
\begin{align}
&E = \frac{16 \pi}{\kappa^2} M, \qquad T = \frac{4 M r_h - q^2 - p^2}{8 \pi r_h^3}, \qquad S = \frac{16 \pi^2 r_h^2}{\kappa^2},\\
&\Phi_{\textrm{E}} = \frac{1}{\kappa} \frac{q}{r_h}, \qquad \Phi_{\textrm{M}} = \frac{1}{\kappa} \frac{p}{r_h}, \qquad Q_E = \frac{4 \pi}{\kappa} q, \qquad Q_M = \frac{4 \pi}{\kappa} p \label{app:Phi_E}
\end{align}
and satisfy the first law of thermodynamics 
\begin{equation}
dE = T dS +\Phi_E dQ_E + \Phi_M dQ_M
\end{equation}
where $Q_E$ and $Q_M$ are independent (i.e. not imposing the regularity condition on $A$). Here $\Phi_{\textrm{E}}$ is obtained by $\Phi_{\textrm{E}}^{(2)} = A_{t}(r_h) - A_{t}(\infty)$. 

Now we calculate $\Phi_{\textrm{E}}$  using   $\Phi_{\textrm{E}}^{(1)} = \frac{\partial E}{\partial Q} |_{r_h, p}$ as in \eqref{eq:emp1}. Taking the variation of $I_{\textrm{ren}}$ with respect to $\beta$ first and then $Q_E$, we obtain
\begin{align}
&\frac{\partial I_{\textrm{ren}}}{\partial \beta} = E - \Phi_{\textrm{E}}^{(1)} Q_E, \\
&\frac{\partial^2 I_{\textrm{ren}}}{\partial Q_E \partial \beta} = \frac{\partial E}{\partial Q_E} - \frac{\partial \Phi_{\textrm{E}}^{(1)}}{\partial Q_E} Q_E - \Phi_{\textrm{E}}^{(1)} =  - \frac{\partial \Phi_{\textrm{E}}^{(1)}}{\partial Q_E} Q_E
\end{align}
where the first law is applied in the second line. This final expression will give us an electric potential as follows
\begin{align}
&\frac{\partial I_{\textrm{ren}}}{\partial \beta} = \frac{\partial M}{\partial \beta} \frac{\partial I_{\textrm{ren}}}{\partial M} = \frac{(r_h-M)^2}{2\pi r_h^2}\frac{\partial I_{\textrm{ren}}}{\partial M} 
  = \frac{2 \pi  \left(4 r_h^2+p^2-q^2\right)}{\kappa ^2 r_h}, \\
&\frac{\partial^2 I_{\textrm{ren}}}{\partial Q_E \partial \beta} = - \frac{1}{\kappa}\frac{q}{r_h} = - \frac{\partial \Phi_{\textrm{E}}^{(1)}}{\partial q} q, \\
&\Phi_{\textrm{E}}^{(1)}= \frac{1}{\kappa} \frac{q}{r_h} 
\end{align}
where the last expression yields the same value of $\Phi_{\textrm{E}}^{(2)}$ in (\ref{app:Phi_E}).

\section{Electric potential for dyonic Taub-NUT-AdS \label{sec:appGP}}
 
We now apply the same logic in appendix \ref{sec:appRNbh} to calculate the electric potential for the dyonic Taub-NUT-AdS solution. Firstly let us start with the same expression for free energy in (\ref{app:I_ren}), which is equivalent to the renormalized action value with $\beta$, then we finally obtain
\begin{equation}
\frac{\partial^2 I_{\textrm{ren}}}{\partial Q \partial \beta} =T \frac{\partial S}{\partial Q}  - \frac{\partial \Phi_{\textrm{E}}}{\partial Q} Q  = - \frac{1}{2G} \frac{(Q (r_b^2 + \chi^2) + 2 i P r_b \chi) r_b}{ (r_b^2 - \chi^2)^2}
\end{equation}
where the last expression have  $Q$ as a factor. This indicates that we need to start with free energy or a renormalized action   having both a magnetic potential and magnetic charge  
\begin{equation}
I_{\textrm{ren}} = \beta F = \beta( E - TS - \Phi_{\textrm{E}} Q - \Phi_{\textrm{M}} P)
\end{equation}
from which the energy is calculated via
\begin{equation*}
\partial_\beta I_{\textrm{ren}}  = E - \Phi_{\textrm{E}} Q - \Phi_{\textrm{M}} P
\end{equation*}
as stated in (\ref{eq:defE}). We then calculate the electric and magnetic potentials via the definition of conjugate variables for charges in (\ref{eq:emp1})
\begin{equation*}
\Phi_{E}^{(1)} = \frac{\partial E}{\partial Q} \bigg|_{r_b,P} - T \frac{\partial S}{\partial Q}\bigg|_{r_b,P}, \qquad \Phi_M^{(1)} = \frac{\partial E}{\partial P} \bigg|_{r_b,Q} - T \frac{\partial S}{\partial Q}\bigg|_{r_b,Q}. 
\end{equation*}
Plugging energy in (\ref{eq:defE}) into (\ref{eq:emp1}) and making a variation with respect to $Q$ and $P$ separately, we obtain
\begin{align}
&\partial_{Q} E = \frac{\partial^2 I_{ren}}{\partial Q \partial \beta}  + \frac{\partial \Phi_\textrm{E}^{(1)}}{\partial Q} Q + \Phi_\textrm{E}^{(1)} +  \frac{\partial \Phi_\textrm{M}^{(1)}}{\partial Q} P, \label{app:FL1} \\
&\partial_{P} E = \frac{\partial^2 I_{ren}}{\partial P \partial \beta}  + \frac{\partial \Phi_\textrm{E}^{(1)}}{\partial P} Q + \Phi_\textrm{M}^{(1)} +  \frac{\partial \Phi_\textrm{M}^{(1)}}{\partial P} P.\label{app:FL2} 
\end{align}
Since the first law is written
\begin{align}
&d_Q E = T d_Q S + \Phi_\textrm{E}^{(1)} d_Q Q + \Phi_\textrm{M}^{(1)} d_Q P =T d_Q S + \Phi_\textrm{E}^{(1)}, \\
&d_P E = T d_P S + \Phi_\textrm{E}^{(1)} d_P Q + \Phi_\textrm{M}^{(1)} d_P P =T d_P S + \Phi_\textrm{M}^{(1)},
\end{align}
(\ref{app:FL1}) and (\ref{app:FL2}) can be changed to
\begin{align}
&T d_Q S = \frac{\partial^2 I_{ren}}{\partial Q \partial \beta} + \frac{\partial \Phi_\textrm{E}^{(1)}}{\partial Q} Q + \frac{\partial \Phi_\textrm{M}^{(1)}}{\partial Q} P, \label{eq:elecp} \\
&T d_P S = \frac{\partial^2 I_{ren}}{\partial P \partial \beta} +\frac{\partial \Phi_\textrm{E}^{(1)}}{\partial P} Q + \frac{\partial \Phi_\textrm{M}^{(1)}}{\partial P} P. \label{eq:magp}
\end{align}
Since we know $T$, $S$, and $I_{ren}$, we can compute $\Phi_\textrm{E}^{(1)}$ and $\Phi_\textrm{M}^{(1)}$ by solving two equations (\ref{eq:elecp}) and (\ref{eq:magp}).

\section{Solution for $r_b$ in terms of other parameters }
\label{app:rbfunc}

From \eqref{eq:hofA}, we set $f(r_b) = 0$, obtaining
\begin{align}
&r_{b,1\mp} = 1 - \sqrt{t2} \chi \mp \sqrt{t1}, \qquad r_{b,2\mp} = 1 + \sqrt{t2} \chi \mp \sqrt{t1},
\end{align}
where 
\begin{align}
&t1 = \frac{48 \chi ^2-1}{\sqrt{t2} \chi ^3}+\frac{t3}{(\sqrt{t4}+ \sqrt{t5})^{1/3}}+\frac{4 (\sqrt{t4} + \sqrt{t5})^{1/3}}{\chi }+192 \chi ^2+\frac{1}{\chi ^2}-32, \\
&t2 =-\frac{2 t3}{(\sqrt{4} + \sqrt{5})^{1/3}}-\frac{8 (\sqrt{4}+\sqrt{5})^{1/3}}{\chi }-96 \chi ^2+48 \left(6 \chi ^2-1\right)+\frac{1}{\chi ^2}+16, \\
&t3=4 \chi  \left(192 \left(3 g^2-1\right) \chi ^2-144 Q^2+576 \chi ^4+7\right),\\
&t4=t5^2-\chi ^6 \left(192 \left(3 g^2-1\right) \chi ^2-144 Q^2+576 \chi ^4+7\right)^3, \\
&t5=27 \left(192 \chi ^5-32 \chi ^3+\chi \right) \left(Q^2-4 g^2 \chi ^2\right)-2 \chi ^3 \left(6912 \chi ^6-3456 \chi ^4+414 \chi ^2-5\right)
\end{align}

\section{Gauge field strength \label{app:F2}}

To obtain free energy, we need to calculate the gauge field strength square $F^2$ in the action in Euclidean space.  The indefinite $r$-integrals of each nonvanishing component of $\int dr \sqrt{g} F^2$ are
\begin{align}
&\int dr \sqrt{g} F_{\tau r}F^{\tau r} = \bigg[ \frac{r(4 i P Q r \chi + (Q^2 -P^2)(r^2 + \chi^2))}{2(r^2 - \chi^2)^2} - \frac{(Q^2 + P^2)}{4\chi} \log \bigg( \frac{r-\chi}{r+\chi} \bigg) \bigg] \sqrt{Y(\theta)^2}, \\
&\int dr \sqrt{g} F_{\theta \phi}F^{\theta \phi} = \bigg[ \frac{r(4 i P Q r \chi + (Q^2 - P^2)(r^2 + \chi^2))}{2(r^2 - \chi^2)^2} + \frac{(Q^2+P^2)}{4\chi} \log \bigg( \frac{r-\chi}{r+\chi} \bigg) \bigg] \frac{\lambda'(\theta)^2}{\sqrt{Y(\theta)^2}}
\end{align}
where the range of $r$ should be taken into account in calculation of the renormalized action.

\section*{Acknowledgements}

MP would like to thank Dongsu Bak for very useful and helpful discussion and acknowledges the hospitality at the Perimeter Institute where part of this work was done. MP was supported by a KIAS Individual Grant (PG062001) at Korea Institute for Advanced Study and by Basic Science Research Program through the National Research Foundation of Korea funded by the Ministry of Education (NRF-2016R1D1A1B03933399). LPZ is supported in part by the U.S. Department of Energy under grant DE-SC0007859.

\bibliographystyle{ieeetr}

\bibliography{Taub-NUT-arXiv}

\begin{thebibliography}{10}

\bibitem{Taub:1950ez}
A.~Taub, ``{Empty space-times admitting a three parameter group of motions},''
  {\em Annals Math.}, vol.~53, pp.~472--490, 1951.

\bibitem{Newman:1963yy}
E.~Newman, L.~Tamburino, and T.~Unti, ``{Empty space generalization of the
  Schwarzschild metric},'' {\em J. Math. Phys.}, vol.~4, p.~915, 1963.

\bibitem{Misner:1963fr}
C.~W. Misner, ``{The Flatter regions of Newman, Unti and Tamburino's
  generalized Schwarzschild space},'' {\em J. Math. Phys.}, vol.~4,
  pp.~924--938, 1963.

\bibitem{Hawking:1998jf}
S.~Hawking and C.~Hunter, ``{Gravitational entropy and global structure},''
  {\em Phys. Rev. D}, vol.~59, p.~044025, 1999.

\bibitem{Hawking:1998ct}
S.~Hawking, C.~Hunter, and D.~N. Page, ``{Nut charge, anti-de Sitter space and
  entropy},'' {\em Phys. Rev. D}, vol.~59, p.~044033, 1999.

\bibitem{Mann:1999pc}
R.~B. Mann, ``{Misner string entropy},'' {\em Phys. Rev. D}, vol.~60,
  p.~104047, 1999.

\bibitem{Taylor:1998fd}
M.~Taylor, ``{Higher dimensional Taub-Bolt solutions and the entropy of
  noncompact manifolds},'' 9 1998.

\bibitem{Clarkson:2002uj}
R.~Clarkson, L.~Fatibene, and R.~B. Mann, ``{Thermodynamics of
  (d+1)-dimensional NUT charged AdS space-times},'' {\em Nucl. Phys. B},
  vol.~652, pp.~348--382, 2003.

\bibitem{AlonsoAlberca:2000cs}
N.~Alonso-Alberca, P.~Meessen, and T.~Ortin, ``{Supersymmetry of topological
  Kerr-Newman-Taub-NUT-AdS space-times},'' {\em Class. Quant. Grav.}, vol.~17,
  pp.~2783--2798, 2000.

\bibitem{Martelli:2012sz}
D.~Martelli, A.~Passias, and J.~Sparks, ``{The supersymmetric NUTs and bolts of
  holography},'' {\em Nucl. Phys. B}, vol.~876, pp.~810--870, 2013.

\bibitem{Toldo:2017qsh}
C.~Toldo and B.~Willett, ``{Partition functions on 3d circle bundles and their
  gravity duals},'' {\em JHEP}, vol.~05, p.~116, 2018.

\bibitem{Clement:2015cxa}
G.~Clément, D.~Gal'tsov, and M.~Guenouche, ``{Rehabilitating space-times with
  NUTs},'' {\em Phys. Lett. B}, vol.~750, pp.~591--594, 2015.

\bibitem{Kubiznak:2019yiu}
R.~A. Hennigar, D.~Kubiz\v{n}ák, and R.~B. Mann, ``{Thermodynamics of
  Lorentzian Taub-NUT spacetimes},'' {\em Phys. Rev. D}, vol.~100, no.~6,
  p.~064055, 2019.

\bibitem{Ballon:2019uha}
A.~B. Bordo, F.~Gray, and D.~Kubiz\v{n}ák, ``{Thermodynamics and Phase
  Transitions of NUTty Dyons},'' {\em JHEP}, vol.~07, p.~119, 2019.

\bibitem{Bordo:2020kxm}
A.~Ballon~Bordo, F.~Gray, and D.~Kubiz\v{n}ák, ``{Thermodynamics of Rotating
  NUTty Dyons},'' {\em JHEP}, vol.~05, p.~084, 2020.

\bibitem{Chamblin:1998pz}
A.~Chamblin, R.~Emparan, C.~V. Johnson, and R.~C. Myers, ``{Large N phases,
  gravitational instantons and the nuts and bolts of AdS holography},'' {\em
  Phys. Rev. D}, vol.~59, p.~064010, 1999.

\bibitem{Johnson:2014xza}
C.~V. Johnson, ``{Thermodynamic Volumes for AdS-Taub-NUT and AdS-Taub-Bolt},''
  {\em Class. Quant. Grav.}, vol.~31, no.~23, p.~235003, 2014.

\bibitem{Johnson:2014pwa}
C.~V. Johnson, ``{The Extended Thermodynamic Phase Structure of Taub-NUT and
  Taub-Bolt},'' {\em Class. Quant. Grav.}, vol.~31, p.~225005, 2014.

\bibitem{Choi:2018vbz}
S.~Choi, J.~Kim, S.~Kim, and J.~Nahmgoong, ``{Comments on deconfinement in
  AdS/CFT},'' 11 2018.

\bibitem{Kubiznak:2016qmn}
D.~Kubiznak, R.~B. Mann, and M.~Teo, ``{Black hole chemistry: thermodynamics
  with Lambda},'' {\em Class. Quant. Grav.}, vol.~34, no.~6, p.~063001, 2017.

\bibitem{Balasubramanian:1999re}
V.~Balasubramanian and P.~Kraus, ``{A Stress tensor for Anti-de Sitter
  gravity},'' {\em Commun. Math. Phys.}, vol.~208, pp.~413--428, 1999.

\bibitem{Emparan:1999pm}
R.~Emparan, C.~V. Johnson, and R.~C. Myers, ``{Surface terms as counterterms in
  the AdS / CFT correspondence},'' {\em Phys. Rev. D}, vol.~60, p.~104001,
  1999.

\bibitem{Mann:2004mi}
R.~B. Mann and C.~Stelea, ``{On the thermodynamics of NUT charged spaces},''
  {\em Phys. Rev. D}, vol.~72, p.~084032, 2005.

\bibitem{Smith:1997wx}
W.~Smith and R.~B. Mann, ``{Formation of topological black holes from
  gravitational collapse},'' {\em Phys. Rev. D}, vol.~56, pp.~4942--4947, 1997.

\bibitem{Lee:2018hrd}
Y.~Lee, M.~Richards, S.~Stotyn, and M.~Park, ``{Quasilocal Smarr relation for
  an asymptotically flat spacetime},'' 9 2018.

\end{thebibliography}

\end{document}